\documentclass[twocolumn,aps,pre,longbibliography,superscriptaddress,10pt]{revtex4-2}
\usepackage{amsmath,amssymb,graphicx,dsfont,float}
\usepackage[colorlinks=true,allcolors=blue]{hyperref}

\begin{document}

\title{Distributional statistics of nonequilibrium}

\author{Harrison Hartle}
\email{hhartle@santafe.edu}
\affiliation{Santa Fe Institute, Santa Fe, New Mexico 87501, USA}

\author{Aanjaneya Kumar}
\affiliation{Santa Fe Institute, Santa Fe, New Mexico 87501, USA}

\author{David H. Wolpert}
\affiliation{Santa Fe Institute, Santa Fe, New Mexico 87501, USA}

\begin{abstract}
We establish a reference modeling framework for random probability distributions subject to constraints. Generalizing the hyperensemble approach of Crooks, we formulate Jaynesian priors of maximal uncertainty over the unit simplex of categorical distributions. This framework is entirely general, but we primarily investigate models with constraints on physically relevant quantities, namely, average energy and Shannon entropy. A constraint on the exact value of average energy uniquely determines a Boltzmann-Gibbs equilibrium, but also defines a hyperplanar subset of the probability simplex consisting of all distributions with that value of average energy. The associated hyperensemble has a uniform probability density across all such distributions, regardless of distance to equilibrium. Such reference models make predictions of generic nonequilibrium properties such as the mean and modal microstate distributions and the expected free energy of a random distribution, which we examine analytically and via exact sampling algorithms. Moreover, introducing constraints on Shannon entropy permits parametric access to tunably nonequilibrium regimes. The approach follows conventional Jaynesian modeling logic and is compatible with standard Bayesian inference, and reduces to Boltzmann-Gibbs statistics when Shannon entropy is maximized. We examine the physical predictions of this scheme, and demonstrate its wide compatibility with other approaches to nonequilibrium modeling. We illustrate the framework in three scenarios: (i) comparison to superstatistics quasi-equilibrium modeling, (ii) generation of random nonequilibrium steady states, and (iii) Bayesian inference under a prior hyperensemble. We highlight numerous prospects for theoretical extensions and empirical application.
\end{abstract}

\maketitle

\tableofcontents

\section{Introduction}
\label{sec:introduction}

Probabilistic modeling in the strictest sense requires assignment of a probability distribution $p$ over outcomes $x$ in a chosen sample space $X$, with $\sum_{x\in X} p_x=1$ and $0\le p_x\le 1$ for each $x\in X$. One may instead opt to remain uncertain about $p$, but at the cost of needing to prescribe a probability density $P(p)$, i.e., a prior, over what $p$ could be. If $X$ is discrete and finite, the distribution $p$ resides in the unit simplex $\Delta(X)$---the set of all categorical distributions over $X$. In statistical mechanics, the condition of {\it equilibrium} sidesteps the need for a nondegenerate $P(p)$, instead representing system statistics by the Boltzmann-Gibbs (BG) distribution $\pi^\beta$. In particular, $\pi^\beta$ maximizes Shannon entropy $S(p)$ among all distributions with a given value of average energy. As shown by Jaynes \cite{Jaynes1957}, $\pi^\beta$ can be seen as the solution to a model selection problem: it is the most {\it unbiased} distribution compatible with a known value of average energy. In particular, this unbiasedness is with respect to the {\it microstate} $x$ drawn from $\pi^\beta$.

However, from the standpoint of an uncertain $p$ as described by a density $P(p)$, the choice of a single $p=\pi^\beta$ amounts to $P(p)=\delta(p-\pi^\beta)$, where $\delta(\cdot)$ denotes the Dirac delta function; any such hyperensemble is maximally {\it biased}, fully concentrated on $\pi^\beta$. Indeed, subject only to the physical constraint of $E(p)=E^*$, the least biased $P(p)$ is uniform across all distributions for which $E(p)=E^*$. Jaynes's formulation can be viewed as implementing the {\it physical} constraint of $S(p)=S_{\mathrm{max}}(E^*)$, which, together with the energetic constraint, uniquely pins down $\pi^\beta$. That is, the constraint $S(p)=S_{\mathrm{max}}(E^*)$ imposes physical equilibrium, which $\pi^\beta$ is indeed an extremely accurate model of. Yet, one may remove this constraint by examining non-maximal or entirely unconstrained $S(p)$, which is the approach we take herein. This requires prescription of a $P(p)$, which in turn makes nonequilibrium physical predictions. We show that even in the absence of Shannon entropy maximization, the logic of Jaynes can be preserved, at the level of $P(p)$ rather than $p$.

More generally, across modeling scenarios involving uncertain probability distributions, the imposed constraints may not determine a unique distribution; yet nevertheless, we show that one can straightforwardly apply Jaynesian model selection: represent incomplete knowledge of $p$ via a probability density of distributions or hyperprior $P(p)$, and moreover, choose the maximally `unbiased' $P(p)$ compatible with knowledge at hand. We consider finite sample spaces $X$ consisting of $n:=|X|$ outcomes, so that $P(p)$ is a density over a finite-dimensional probability simplex $\Delta(X)$. The relevant uncertainty to maximize is the differential entropy $\mathbb{S}(P)$ or more specifically, the negative Kullback-Leibler divergence between $P$ and a background measure $M$:
\begin{equation}
\label{eq:S_of_P}
\mathbb{S}(P)=-\int_{\Delta(X)}P(p)\log\frac{P(p)}{M(p)}dp,
\end{equation}
with $\int_{\Delta(X)}dp:=\int_0^1\int_0^{1-p_1}\cdots \int_0^{1-\sum_{i=1}^{n-2}p_i}\prod_{i=1}^{n-1}dp_{i}$ and $p_n:=1-\sum_{i=1}^{n-1}p_i$ in any integrand argument. We for simplicity consider herein the uniform $M(p)=\mathrm{const}$ as the background measure (in particular, $M(p)=1/|\Delta(X)|=(n-1)!$), but the Jeffreys prior would also be a natural choice \cite{Jeffreys}. The hyperensemble $P(p)$ is taken as that which maximizes differential entropy $\mathbb{S}(P)$ subject to imposed constraints. The obtained $P(p)$ can be treated as a model of population-level distributional statistics, choosing constraints to reproduce some properties of datasets consisting of distributions; or, $P(p)$ can be treated as a Bayesian prior over categorical distributions, suitable for inference. Equivalently, $P(p)$ is a generalized statistical mechanics model~\cite{davis2016} with its microstates being probability distributions~\cite{Balasubramanian1997}, also amenable to field-theoretic interpretations~\cite{Bialek1996}. Shannon entropy $S(p)$ can be incorporated as a constraint, and when maximized, traditional maximum entropy models over $X$ are recovered~\cite{Presse2013}.

We demonstrate the straightforward applicability of this hyperstatistical picture to nonequilibrium physical modeling. We consider first the simplest scenarios with only energetic constraints in Sec.~\ref{ssec:energetic_hyperensembles}, and then incorporate entropic constraints in Sec.~\ref{sec:entropic_constraint}. We illustrate these hyperensembles in Sec.~\ref{ssec:level_spacing} for a special-case Hamiltonian, examining the mean and modal distributions, and the statistics of nonequilibrium free energy, in relation to the associated BG equilibria. In Sec.~\ref{sec:applications}, we formulate applications in a few settings: superstatistics, nonequilibrium steady states, and Bayesian inference. In Sec.~\ref{sec:discussion} we provide concluding discussion. Appendices~\ref{app:entropy},~\ref{app:dirichlet_entropic_crooks},~\ref{app:hitandrun},~\ref{app:min_ent},~\ref{app:Zgamma},~\ref{app:rej_mcmc},~\ref{app:beta0_KLvals},~\ref{app:f_beta_for_P_gamma}, and~\ref{app:mode_F} cover supporting details.

\section{Framework}
\label{sec:framework}

Here we spell out a general framework for formulating models of random probability distributions, a.k.a., hyperensembles, denoted $P(p)$, subject to constraints on outcomes $p\sim P$. We consider discrete sample spaces $X$ with $n:=|X|$ finite. In the absence of any constraints whatsoever, the $\mathbb{S}$-maximizer is $\mathrm{Dirichlet}(\mathbf{1})$, i.e., the uniform prior $\mathrm{U}[\Delta(X)]$, with probability density $P(p)=|\Delta(X)|^{-1}=(n-1)!$, representing perfect ignorance of the distribution $p\in \Delta(X)$. By the null-modeling logic of Jaynes, one's `knowledge' can be encoded by imposing constraints on the model, then among the remaining constraint-compatible models, the uncertainty-maximizing one is selected. Here, constraints pertain to properties of $p$, and the resulting $P(p)$ is chosen to maximize $\mathbb{S}(P)$. We consider two forms of constraint---exact constraints and expectation constraints---which, respectively, are akin to the microcanonical and canonical constraints of equilibrium statistical mechanics~\cite{Kardar_2007}; these are the standard options in maximum-entropy modeling~\cite{squartini2017maximum}. The sample space consists of the unit simplex $\Delta(X)$ of categorical distributions on $X$, requiring these constraints to be expressed in terms of functions of $p$. The two forms of constraint are outlined below:
\ \\

{\it Exact constraints}: these require that every distribution $p\sim P$ has the prescribed value of a given function $Q(p)$, yielding a subset of $\Delta(X)$ onto which the support of $P$ is restricted. Namely, 
\begin{equation}
\label{eq:Qk}
Q(p)=Q^*,
\end{equation}
restricting the support of $P$ to the simplex-subset 
\begin{equation}
\Delta_{Q^*}:=\{p\in\Delta(X):Q(p)=Q^*\}.
\end{equation}
The dimensionality of $\Delta_{Q^*}$ may be lower than that of $\Delta(X)$, in which case integrals such as Eq.\,\eqref{eq:S_of_P} have the associated surface element as $dp$. Inequality constraints can be incorporated by choosing $Q(p)$ as an appropriate indicator function~\cite{Kazama_2005}. We note that acquiring perfect knowledge of the constraint-value $Q^*$ is an idealization, but a widely adopted one in maximum-entropy modeling~\cite{Kapur_1992,De_Martino_2018,Golan_2022}; a variety of approaches have been developed to estimate functions of unknown distributions~\cite{Jiao_2017}.\\

{\it Expectation constraints:} these require that the expectation of a given function $F(p)$ for $p\sim P$ has a prescribed value:
\begin{equation}
\label{eq:Fj}
\langle F(p)\rangle_P=\overline{F},
\end{equation}
where the hyperensemble expectation is denoted
\begin{equation}
\langle F(p)\rangle_P:=\int_{\Delta(X)}P(p)F(p)dp.
\end{equation}

The probability density $P(p)$ that maximizes $\mathbb{S}(P)$ subject to Eq.\,\eqref{eq:Fj} is expressed in terms of a Lagrange multiplier (or vector thereof) $\lambda$ with value(s) determined by $\overline{F}$. Denoting $1_{Q^*}(p):=\mathds{1}\{Q(p)=Q^*\}$ as the indicator enforcing Eq.\,\eqref{eq:Qk}, the hyperensemble subject to both Eq.\,\eqref{eq:Qk} and Eq.\,\eqref{eq:Fj} and maximizing Eq.\,\eqref{eq:S_of_P} under the uniform background measure has density $P(p)$ given by 
\begin{equation}
P_{Q^*,\lambda}(p)=\frac{1_{Q^*}(p)e^{\lambda F(p)}}{Z(Q^*,\lambda)},
\end{equation}
with partition function $Z(Q^*,\lambda)$ given by
\begin{equation}
Z(Q^*,\lambda)=\int_{\Delta_{Q^*}}e^{\lambda F(p)}dp,
\end{equation}
and the value of $\overline{F}$ is related to $\lambda$ by 
\begin{equation}
\label{eq:dlogZdlambda} 
\overline{F}=\frac{\partial}{\partial\lambda}\log Z(Q^*,\lambda),
\end{equation}
see Appendix~\ref{app:entropy}. This framework is compatible with Bayesian inference, in that any hyperensemble $P(p)$ can be treated as a prior over the set of categorical distributions $\Delta(X)$, and defines a marginal over microstates, $\overline{p}=\int_{\Delta(X)} pP(p)dp$, the mean distribution: $\overline{p}_x$ the prior probability of $x$. Observation of $x$ then implies, by Bayes's rule, a posterior hyperensemble $P(p|x)=P(p)p_x/\overline{p}_x$, with $p_x$ the conditional probability or likelihood of $x$ given $p$. These considerations apply also to multiple observations (see Sec.~\ref{ssec:inference}). Moreover, one may include uncertainty in $\overline{F}$ or $Q^*$, permitting Bayesian treatments at the level of $P$ rather than just $p$~\cite{Kang_2021}.

\subsection{Scope of application}
\label{ssec:scope}

The framework outlined above can be applied whenever the available information constrains a family of possible probability distributions $p$ over a finite set of outcomes $X$ without selecting a unique one. In such cases, the relevant modeling problem is not only to assign probabilities to outcomes $x\in X$, but to assign a probability density $P(p)$ over the possible laws governing those outcomes. The hyperensemble $P(p)$ is then a null model for incomplete knowledge of $p$: it is the least biased ($\mathbb{S}$-maximizing) density over $\Delta(X)$ compatible with the imposed constraints. Data in the form of probability distributions are commonly encountered---i.e., normalized histograms, relative abundances, or compositional data~\cite{aitchison1982,BaconShone_1992,morrison2025}; indeed, the Jaynesian approach to compositional modeling has been explored in Ref.~\cite{Weistuch_2022} under constraints on $\langle p_x\rangle$ and $\langle p_x p_{x'}\rangle$. See Appendix~\ref{app:dirichlet_entropic_crooks} for background on other related past works. Further examples of settings to which this approach is naturally applicable include the following.

\begin{figure*}
    \centering
    \includegraphics[width=\linewidth,trim=0 60 0 20]{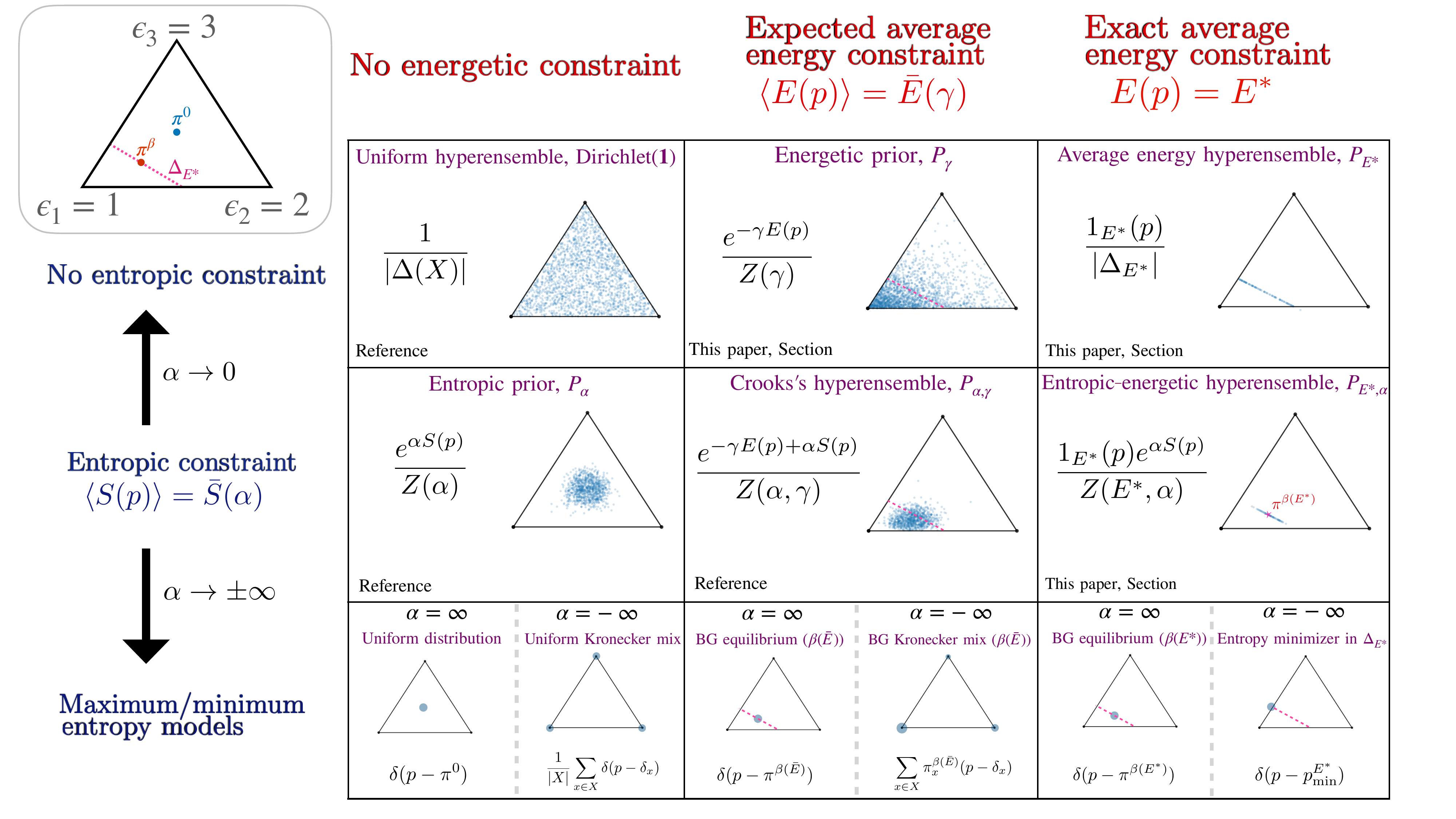}
    \begin{picture}(0,0)
                \put(161,159){\makebox(0,0)[l]{{\scriptsize\ref{ssec:exact_average_energy}}}}
                \put(39,159.5){\makebox(0,0)[l]{{\scriptsize\ref{ssec:expected_average_energy}}}}   
                \put(159,77){\makebox(0,0)[l]{{\scriptsize\ref{ssec:entropic_energetic}}}}
                \put(-104,159){\makebox(20,0)[l]{{\scriptsize\cite{Dirichlet_2012}}}}
                \put(-102,77){\makebox(0,0)[l]{{\scriptsize\cite{caticha2004entropic}}}}
                \put(15,77){\makebox(20,0)[l]{{\scriptsize\cite{crooks2007beyond}}}}
    \end{picture}
    \caption{Hyperensembles that maximize $\mathbb{S}(P)$ subject to constraints on average energy $E(p)$ and Shannon entropy $S(p)$ of $p\sim P$. Each panel depicts the form of $P(p)$ alongside samples $p\sim P$ in the unit simplex of distributions over $X=\{1,2,3\}$ with energies $\epsilon_x=x$. We consider two forms of constraint on average energy $E(p)$, either setting its hyperensemble expectation to $\overline{E}$ (middle column), or setting its exact value to $E^*$ (right column). We also consider an expectation constraint on Shannon entropy (middle row) yielding $P(p)\propto e^{\alpha S(p)}$ with Lagrange multiplier $\alpha$; this includes the entropic prior $P_\alpha$, Crooks's hyperensemble $P_{\alpha,\gamma}$ and the entropic-energetic hyperensemble $P_{E^*,\alpha}$. The limit $\alpha=0$ (upper row) yields entropically unconstrained models, whereas $\alpha\rightarrow\pm\infty$ (lower row) yields hyperensembles concentrated on distributions with extremal values of entropy. In particular, at $\alpha=\infty$, the support consists of the unique Boltzmann-Gibbs equilibrium; at $\alpha=-\infty$, the support consists of the entropy minimizer(s), yielding a mixture of Kronecker deltas in $P_\alpha$ and $P_{\alpha,\gamma}$ (achieving $\overline{S}=0$), and a minimizer $p_{E^*}^{\mathrm{min}}$ with nonvanishing entropy in the case of $P_{E^*,\alpha}$. The models $\mathrm{Dirichlet}(\mathbf{1})$, $P_\alpha$, and $P_{\alpha,\gamma}$ were introduced previously~\cite{Dirichlet_2012,caticha2004entropic,crooks2007beyond} (see Appendix~\ref{app:dirichlet_entropic_crooks}) and are generalized by the framework of Sec.~\ref{sec:framework}. We introduce the models $P_{E^*}$, $P_{\gamma}$, and $P_{E^*,\alpha}$ in Sec.~\ref{sec:therm_hyp}.}
    \label{fig:hyp_ens_schem}
\end{figure*}

\textit{Statistical mechanics.} Here $x$ denotes a microstate and $p_x$ its probability. Constraints on $p$ may involve average energy, Shannon entropy, free energy, particle number, or other thermodynamic quantities. $P(p)$ represents epistemic uncertainty of the probabilistic state $p$, or a physically present `ensemble of ensembles' arising from layered stochasticity. The statistical-mechanical setting is our primary focus herein, with thermodynamic hyperensembles introduced in Sec.~\ref{sec:therm_hyp} and applications considered in Sec.~\ref{sec:applications}. This approach to physical modeling was introduced by Crooks~\cite{crooks2007beyond} and has subsequently been applied to estimate equilibration rates~\cite{Gao2016} and for prebiotic chemistry~\cite{Stopnitzky2019}. For an overview of the physical constraints and hyperensembles we consider, see Fig.~\ref{fig:hyp_ens_schem}.

\textit{Single-cell or population biology.} One may consider $x$ to denote a species, genotype, phenotype, cellular state, or mutation, while $p_x$ denotes its abundance or frequency in a population~\cite{Prentice_1986,Gloor_2017,Hart_2026}. Across environments, individuals, patients, tissues, or experimental preparations, $p$ may vary~\cite{Aebischer_1993,Nichols_2023,morrison2022,Morrison2023}. A hyperensemble $P(p)$ then represents variability over a collection of population compositions, or epistemic uncertainty about a particular population composition, subject to constraints such as aggregate abundances, biodiversity, environmental variables, or conserved resources.

\textit{Social systems and data science.} In these applications, $x$ can represent an individual's state of mind, opinion, communications, behavior, wealth, demographics, or health status; a given distribution $p$ can describe population-level proportions of these variables, or their variation in an individual throughout a relevant timescale~\cite{Chastin_2015,Kjaergaard2019}. Analogous quantities of interest in social systems are defined at household-level, organizational, or governmental scales~\cite{Shelton2007,Fry2000,JofreCampuzano2022}. A hyperensemble $P(p)$ then describes how $p$ varies throughout a metapopulation, or one's epistemic uncertainty in $p$. Constraints may involve material resources, time, mean activity, inequality, or conservation laws.

\textit{Machine learning.} Here $x$ may denote a datapoint in a training set, including its features and label, and $p$ the data-generating distribution in a deployment environment. When a model is trained in one setting and used in another, the distribution of data may change in ways that are only partially known~\cite{Koh2021,QuinoneroCandela2008}. A hyperensemble $P(p)$ represents uncertainty over the distributions that might be encountered in use. Constraints may specify how often each category appears, average values of selected features, or a limit on how much the data distribution can differ from that used for training. More generally, numerous distributions arising in machine learning are amenable to hyperstatistical extensions~\cite{Vapnik_1999,murphy2012machine}.

This handful of potential application settings, and the mentioned forms of distribution and hyperensemble for each, is very far from exhaustive. However, it is indicative of the wide generality and utility of a Jaynesian modeling framework for random probability distributions. In the remainder of this article, we focus on applications in statistical mechanics, and in particular, for the modeling of nonequilibrium physical systems. We define and impose constraints of physical interest, and analyze the nonequilibrium distributional statistics predicted by the associated $\mathbb{S}$-maximizing hyperensembles. Fig.~\ref{fig:hyp_ens_schem} depicts constraint choices and resulting hyperensembles.

\section{Thermodynamic hyperensembles}
\label{sec:therm_hyp}

We introduce three hyperensembles of thermodynamic relevance, defined by constraints on energy (Sec.~\ref{ssec:energetic_hyperensembles}), and Shannon entropy (Sec.~\ref{sec:entropic_constraint}), comparing them to Boltzmann-Gibbs statistics in Sec.~\ref{ssec:level_spacing}.

\subsection{Energetic constraints}
\label{ssec:energetic_hyperensembles}

We begin with consideration of hyperensembles subject to a constraint on the average energy $E(p)$, which is defined as
\begin{equation}
\label{eq:E_def}
E(p):=\sum_{x\in X}p_x\epsilon_x,
\end{equation}
with energies $\{\epsilon_x\}_{x\in X}$ assumed to be known and distinct. We denote $\epsilon_+:=\max_{x\in X}\epsilon_x$ and $\epsilon_{-}:=\min_{x\in X}\epsilon_x$, and $x_{\pm}\in X$ as the corresponding microstates ($\epsilon_{x_{\pm}}=\epsilon_{\pm}$). We also denote $\mathcal{E}:=|X|^{-1}\sum_{x\in X}\epsilon_x$ as the arithmetic mean energy. The Kronecker delta distribution at microstate $x$ is denoted $\delta_x:=(\delta_{x,x'})_{x'\in X}$.

In what follows, we present a variety of analytical and algorithmic results for two energy-constrained hyperensembles formulated as per Sec.~\ref{sec:framework}. Namely, we first consider an exact constraint on average energy,
\begin{equation}
\label{eq:Estar_constrant}
E(p)=E^*,
\end{equation}
 resulting in $P_{E^*}(p)\propto \mathds{1}\{E(p)=E^*\}$ (Sec.~\ref{ssec:exact_average_energy}). We also consider an expected average energy constraint,
\begin{equation}
\label{eq:barE_constrant}
\langle E(p)\rangle=\overline{E}.
\end{equation}
resulting in $P_\gamma(p)\propto e^{-\gamma E(p)}$ (Sec.~\ref{ssec:expected_average_energy}). For reference, we compare to the associated Boltzmann-Gibbs (BG) equilibrium $\pi^\beta$ under parameter-matchings derived in Sec.~\ref{sssec:param_matchings}. These models capture how far from equilibrium one would generically expect a system to be, given knowledge only of energetics in the form of constraints on $E(p)$. Models that additionally incorporate constraints on entropy, and hence exhibit tunably nonequilibrium statistics, are considered in Sec.~\ref{sec:entropic_constraint}.

\subsubsection{Average energy hyperensemble}
\label{ssec:exact_average_energy}

Here we introduce the average energy hyperensemble, a uniform probability density $\mathrm{U}[\Delta_{E^*}]$ over
\begin{equation}
\Delta_{E^*}:=\{p\in\Delta(X):E(p)=E^*\},
\end{equation}
representing an unbiased hyperensemble subject to an exact constraint on the average energy. This model has density $P_{E^*}(p)=\mathds{1}\{E(p)=E^*\}/|\Delta_{E^*}|$, with $|\Delta_{E^*}|:=\int_{\Delta_{E^*}}dp$ the $(n-2)$-dimensional volume of $\Delta_{E^*}$.

Jaynes's formulation of the canonical (BG) ensemble, $\pi^\beta$, demonstrates that $\pi^\beta$ is the `most unbiased' model for which $E(p)=E^*$, in that across all distributions in $\Delta_{E^*}$, the maximizer of Shannon entropy is $\pi^\beta$. But from the hyperstatistical standpoint and framework of Sec.~\ref{sec:framework}, imposition of $S(p)=S_{\mathrm{max}}(E^*)$ would constitute a second constraint, and moreover a physical assumption. Instead, maximization of $\mathbb{S}$ yields $P_{E^*}$ as the `most unbiased' model. In equilibrium, specification of $E^*$ implies a unique inverse temperature $\beta$ that can be associated with an infinite thermal bath. For $P_{E^*}$, a notion of `energy balance' can still apply, but without thermalization to equilibrium; the distributional randomness of $P_{E^*}$ can be realized by a `heatless' stochastic algorithm that traverses $\Delta_{E^*}$, which we turn to next.

\begin{figure}[H]
    \centering
    \includegraphics[width=\linewidth,trim=0 20 0 20]{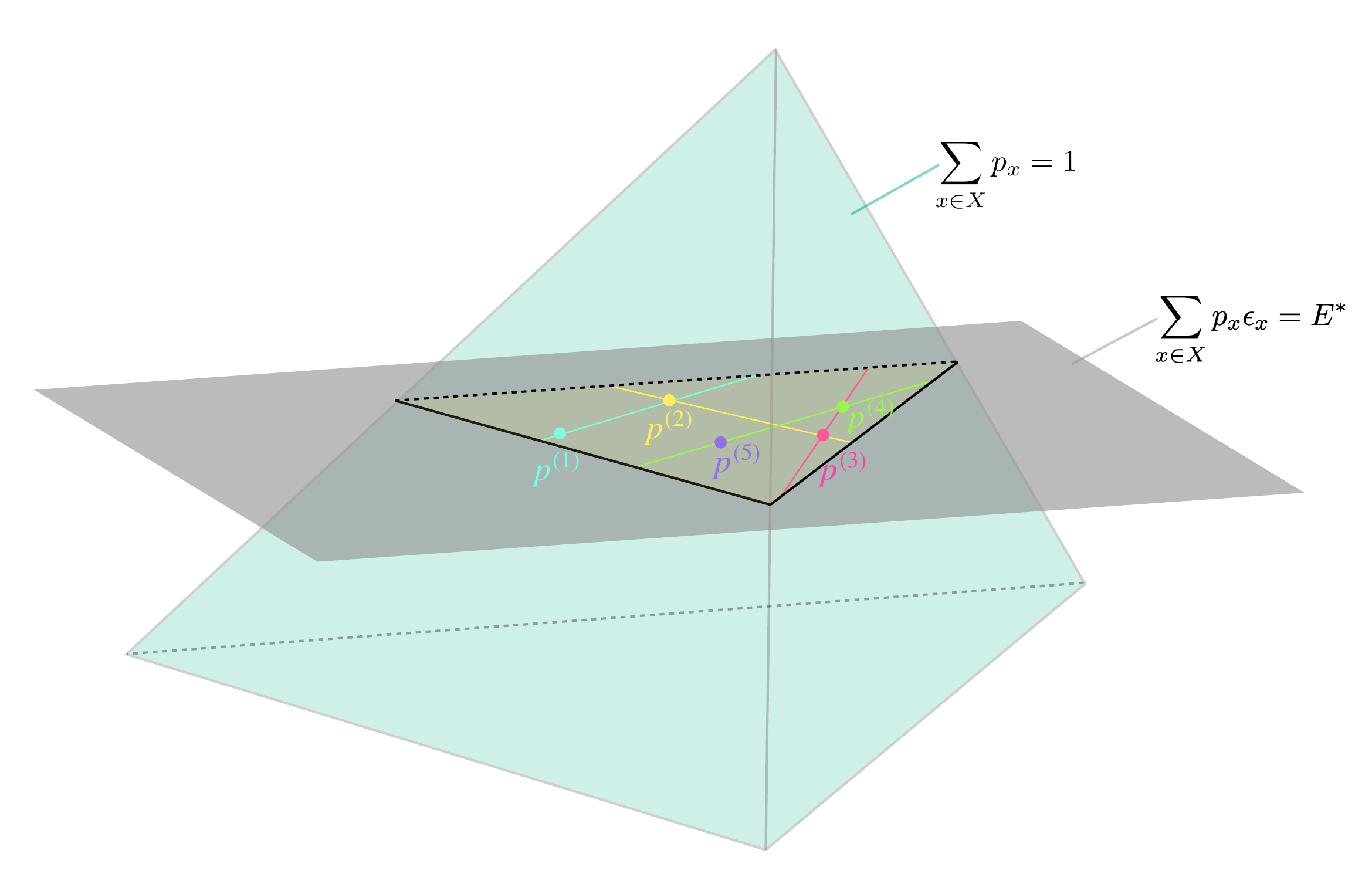}
    \caption{The hit-and-run algorithm $\mathbf{HR}(\Delta_{E^*})$ to sample hyperensemble $P_{E^*}$; details are covered in Appendix~\ref{app:hitandrun}.}\label{fig:hitandrunschematic}
\end{figure}

The level sets of $E(p)$ are hyperplanar, by its linearity (Eq.\,~\eqref{eq:E_def}). The unit simplex $\Delta(X)$ is a convex polytope, so its intersection $\Delta_{E^*}$ with the hyperplane $E(p)=E^*$ is also a convex polytope. Therefore, the average energy hyperensemble $P_{E^*}$ can be sampled via the hit-and-run algorithm \cite{hitandrun1993}. This algorithm is denoted $\mathbf{HR}(\Delta_{E^*})$ and takes the form:
 \begin{enumerate}
     \item Initialize at arbitrary $p^{(1)}\in\Delta_{E^*}$ at index $t=1$.
     \item Sample a vector $u$, randomly-oriented within $\Delta_{E^*}$.
     \item Draw a line segment through $p^{(t)}$ parallel to $u$.
     \item Sample $p^{(t+1)}$ uniform-randomly in that segment.
     \item Repeat 2.--\,4. at $t=2,\ldots,T-1$ for a chosen $T$.
\end{enumerate} 
See Fig.~\ref{fig:hitandrunschematic} for an illustration of $\mathbf{HR}(\Delta_{E^*})$. 

Hit-and-run has rapid mixing \cite{lovasz1999hit}, with the number of steps $T$ to produce a uniform sample being of the order $|X|^3$, up to logarithmic factors. Appendix~\ref{app:hitandrun} covers $\mathbf{HR}(\Delta_{E^*})$ in more depth, including how $p^{(1)}$ is constructed, how to sample a randomly orientated vector $u$ while preserving normalization and average energy, and how to obtain the line segment's intersections with boundaries of $\Delta_{E^*}$. Simple extensions also allows efficient sampling from hyperensembles subject to multiple exact or expectation constraints.

The mode of $P_{E^*}$ is not unique, with all distributions in $\Delta_{E^*}$ having the same probability density. The average distribution $\overline{p}$ resides within $\Delta_{E^*}$, and has a geometric interpretation as the centroid:
 \begin{equation}
 \label{eq:barp_Estar}
\overline{p}:=\langle p\rangle_{ P_{E^*}}=\frac{\int_{\Delta_{E^*}}pdp}{|\Delta_{E^*}|}=\mathrm{centroid}(\Delta_{E^*}),
\end{equation}
which can be computed As another characterization of $\Delta_{E^*}$ and $P_{E^*}$, we examine the range of Shannon entropies $S(p)$ among $p\in \Delta_{E^*}$. That range determines the support of $S(p)$ for $p\sim P_{E^*}$, and is relevant to Sec.~\ref{sec:entropic_constraint} in which entropic constraints are incorporated. The maximum value is $S_{\mathrm{max}}(E^*)=S(\pi^\beta)=\beta E^*+\log Z_\beta$ by the definition of $\pi^\beta$. The minimum value is achieved at an outermost corner vertex of $\Delta_{E^*}$ by a distribution $p^{\mathrm{min}}_{E^*}$ consisting of a mixture of two Kronecker deltas at microstates with energies above and below $E^*$. The minimal value of entropy is thus upper-bounded as $S_{\mathrm{min}}(E^*)\le \log 2$. In particular, denoting $h(\cdot)$ for the binary entropy function, we show in Appendix~\ref{app:min_ent} that 
\begin{equation}
\label{eq:min_pair}
S_{\mathrm{min}}(E^*)=\min\left(h\left(\frac{\epsilon_>-E^*}{\epsilon_>-\epsilon_-}\right),h\left(\frac{E^*-\epsilon_<}{\epsilon_{+}-\epsilon_{<}}\right)\right)
\end{equation}
with $\epsilon_{>}$ and $\epsilon_{<}$ denoting the smallest energy above $E^*$ and largest energy below $E^*$, respectively. If $E^*=\epsilon_{x'}$ for some $x'\in X$, then $p^{\mathrm{min}}_{E^*}=\delta_{x'}$, attaining $S_{\mathrm{min}}(E^*)=0$.

\subsubsection{Expected average energy hyperensemble}
\label{ssec:expected_average_energy}

We also consider expectation constraints on average energy $E(p)$, as per Eq.\,\eqref{eq:Fj}, of the form $\langle E(p)\rangle=\overline{E}$. The resulting $\mathbb{S}$-maximizing hyperensemble has a Lagrange multiplier denoted $\gamma$ and takes the form
\begin{equation}
\label{eq:P_gamma}
P_{\gamma}(p)=\frac{e^{-\gamma E(p)}}{Z(\gamma)}, \ \ \ \ \ Z(\gamma)=\int_{\Delta(X)}e^{-\gamma E(p)}dp.
\end{equation}
The sign convention of $-\gamma E(p)$ is chosen to mimic $-\beta\epsilon_x$ of the canonical ensemble $\pi^\beta$. By analogy to the entropic prior~\cite{caticha2004entropic}, we refer to $P_{\gamma}$ as the `energetic prior'. The choice of $\overline{E}$ determines $\gamma$, with limiting behavior $\lim_{\gamma\rightarrow\pm \infty}\overline{E}(\gamma)=\epsilon_{\mp}$. In particular, from~\eqref{eq:dlogZdlambda}, we have $\overline{E}=-\frac{d}{d\gamma}\log Z(\gamma)$, which is generally transcendental, but with a unique and numerically tractable root. In Appendix~\ref{app:Zgamma} we apply the result of Ref.~\cite{Barvinok1992} to obtain the general formula 
\begin{equation}
\label{Z_gamma}
Z(\gamma)=\frac{1}{(-\gamma)^{n-1}}\sum_{x\in X} \frac{e^{-\gamma\epsilon_x}}{\prod_{x'\in X\backslash\{x\}}(\epsilon_x -\epsilon_{x'})}.
\end{equation}
The simplest way to draw $p\sim P_\gamma$ is via rejection sampling, accepting a candidate distribution $p'\sim \mathrm{U}[\Delta(X)]$ with probability $\mathcal{A}_\gamma(p')\propto e^{-\gamma E(p)}$; more generally for an MCMC algorithm with $\mathrm{U}[\Delta(X)]$ as its stationary distribution, the acceptance probability is $\mathcal{A}_\gamma(p\rightarrow p')=\min(1,e^{-\gamma (E(p')-E(p))})$ for candidate transition $p\rightarrow p'$ (see Appendix~\ref{app:rej_mcmc}).

Note that if we were to construct a hyperensemble with constraints on the average probability $\langle p_x\rangle=\overline{p}_x$ of each $x\in X$, i.e., to impose a set of $|X|$ distinct expectation constraints with individualized Lagrange multipliers $\boldsymbol{\lambda}=(\lambda_x)_{x\in X}\in\mathbb{R}^{|X|}$, we would arrive at a hyperensemble of density $\propto \exp(\sum_{x\in X}\lambda_x p_x)/Z(\boldsymbol{\lambda})$. That has the same form as $P_\gamma$ (Eq.\,\eqref{eq:P_gamma}) under substitution $\boldsymbol{\lambda}=-\gamma\boldsymbol{\epsilon}$, with $\boldsymbol{\epsilon}:=(\epsilon_x)_{x\in X}$. Accordingly, our choice of {\it Hamiltonian} or the energy values $\boldsymbol{\epsilon}$ alongside $\gamma$ parameterize $\overline{p}=(\overline{p}_x)_{x\in X}$. As such,
\begin{equation}
-\frac{1}{\gamma}\frac{\partial \log Z(\gamma)}{\partial\epsilon_x}
=\frac{1}{ Z(\gamma)}\int_{\Delta(X)}e^{-\gamma\epsilon_x}p_xdp=\overline{p}_x,
\end{equation}
or more concisely, $\overline{p}=-\gamma^{-1}\nabla_{\boldsymbol{\epsilon}}\log Z(\gamma)$. The mode of $P_{\gamma}$ is $\delta_{x_-}$ when $\gamma>0$ and $\delta_{x_+}$ when $\gamma<0$.

\subsection{Entropic constraint}
\label{sec:entropic_constraint}

We constructed hyperensembles $P_{E^*}$ and $P_{\gamma}$ by maximization of $\mathbb{S}$ subject to distributional constraints on average energy $E(p)$. For statistical mechanics these are especially relevant because they impose no bias towards equilibrium. But one may also have knowledge of {\it how nonequilibrium} a given system is~\cite{Vinck2012,Paninski2003,Archer2014}, motivating hyperensembles with constraints on Shannon entropy $S(p):=-\sum_{x\in X}p_x \log p_x$~\cite{shannon}. We consider an expectation constraint, as per~\eqref{eq:Fj}, of the form
\begin{equation}
\label{eq:expected_entropy_constraint}
\langle S(p)\rangle=\overline{S}.
\end{equation}
Without any other constraints, maximization of $\mathbb{S}$ subject to~\eqref{eq:expected_entropy_constraint} yields the entropic prior $P_\alpha(p)$, arising in Bayesian statistics and entropy estimation \cite{brewer2009entropic,Neumann2007,caticha2004entropic,Caticha2004Maximum}, which takes the form
\begin{equation}
P_\alpha(p)=\frac{e^{\alpha S(p)}}{Z(\alpha)}.
\end{equation}
Crooks introduced hyperensembles for which~\eqref{eq:expected_entropy_constraint} is interpretable as a {\it physical} constraint, subject also to $\langle E(p)\rangle=\overline{E}$. Together with~\eqref{eq:expected_entropy_constraint}, the $\mathbb{S}$-maximizer is Crooks's hyperensemble, $P_{\gamma,\alpha}$, which can be written in the form
\begin{equation}
\label{eq:P_crooks_F}
P_{\gamma,\alpha}(p)=\frac{e^{-\gamma F_{\gamma/\alpha}(p)}}{\tilde{Z}(\alpha,\gamma)},
\end{equation}
with free energy $F_\beta(p):=E(p)-\beta^{-1}S(p)$ and $\beta=\gamma/\alpha$ under the parameterization in~\eqref{eq:P_gamma} for $P_\gamma$. All results on $P_{\gamma}$ are recovered in the limit $\alpha=0$.

By generalizing to arbitrary combinations of exact and expectation constraints on $p\sim P$, the framework of Sec.~\ref{sec:framework} includes both $P_\alpha$ and $P_{\gamma,\alpha}$ (see Appendix~\ref{app:dirichlet_entropic_crooks}). We next return to consideration of the exact average energy constraint $E(p)=E^*$, but in conjunction with the expected entropy constraint~\eqref{eq:expected_entropy_constraint}.

\subsubsection{Entropic-energetic hyperensemble}
\label{ssec:entropic_energetic}

Subject to expected entropy constraint~\eqref{eq:expected_entropy_constraint} alongside $E(p)=E^*$, the $\mathbb{S}$-maximizing hyperensemble $P_{E^*,\alpha}$, whose support is $\Delta_{E^*}$, takes the form
\begin{equation}
P_{E^*,\alpha}(p)=\frac{1_{E^*}(p)e^{\alpha S(p)}}{Z(E^*,\alpha)},
\end{equation}
with $Z(E^*,\alpha)=\int_{\Delta_{E^*}}e^{\alpha S(p)}dp$; the parameter $\alpha$ is determined by~\eqref{eq:dlogZdlambda}, i.e., for a prescribed $\overline{S}$, as the solution of
\begin{equation}
\overline{S}=\frac{\partial}{\partial \alpha}\log Z(E^*,\alpha).
\end{equation}
Unlike in $P_{E^*}$ and $P_{\gamma}$, the mean excess free energy becomes a directly tunable property in $P_{E^*,\alpha}$ in terms of constraint values $\overline{S}$ and $E^*$:
\begin{equation}
\label{eq:mean_F_Estar_alpha}
\langle\Delta F_\beta(p)\rangle=E^*-\beta^{-1}\overline{S}+\beta^{-1}\log Z_{\beta},
\end{equation}
with $\beta=\beta(E^*)$ fixed by $E(\pi^{\beta})=-\frac{d}{d\beta}\log Z_\beta=E^*$. Moreover, akin to in $P_{\gamma,\alpha}$ (Appendix~\ref{app:crooks}), we may write
\begin{equation}
\label{eq:P_Estar_alpha_KL}
P_{E^*,\alpha}(p)\propto 1_{E^*}(p)e^{-\alpha D(p||\pi^\beta)}\propto 1_{E^*}(p)e^{-\alpha\beta \Delta F_\beta(p)}.
\end{equation}
Thus, $P_{E^*,\alpha}$ may be interpreted equivalently as the $\mathbb{S}$-maximizing hyperensemble subject to $E(p)=E^*$ with a given expected KL divergence to equilibrium, or a given mean extractable work~\cite{sivak2012nearequilibrium} (see Sec.~\ref{ssec:BG_comparison}).

Neither $P_\alpha$ nor $P_{\gamma,\alpha}$ are recoverable as special cases of $P_{E^*,\alpha}$, because the support of $P_{E^*,\alpha}$ is restricted to $\Delta_{E^*}$. The parameter range for $E^*$ is $[\epsilon_-,\epsilon_+]$, and at a given $E^*$, the parameter range for $\overline{S}$ is $[S_{\mathrm{min}}(E^*),S_{\mathrm{max}}(E^*)]$, with $S_{\mathrm{max}}(E^*)=S(\pi^{\beta(E^*)})$ and with $S_{\mathrm{min}}(E^*)$ given by Eq.\,\eqref{eq:min_pair}. This range for $\overline{S}$ coincides also with the support of $S(p)$ for random $p\sim P_{E^*,\alpha}$, in notable contrast with $p\sim P_{\gamma,\alpha}$: at a given value of $\overline{E}$, the latter permits any $\overline{S}\in[0,S_{\mathrm{max}}(\overline{E})]$, and has a density of $S(p)$ with support on $[0,\log n]$ reflecting that $P$ has support $\Delta(X)$.

For $\alpha>0$, higher-entropy distributions are favored in $P_{E^*,\alpha}$, with limiting extreme $P_{E^*,\infty}=\delta(p-\pi^{\beta(E^*)})$. For $\alpha<0$, {\it low}-entropy distributions are favored. The $\alpha\rightarrow-\infty$ limit corresponds to $\overline{S}\downarrow S_{\mathrm{min}}(E^*)$ (Eq.\,\eqref{eq:min_pair}) with $P_{E^*,\alpha}$ concentrating on the entropy minimizer $p^{\mathrm{min}}_{E^*}$ (Appendix~\ref{app:min_ent}). At $\alpha=0$, the hyperensemble $P_{E^*}$ of Sec.~\ref{ssec:exact_average_energy} is recovered.

\subsection{Comparison to Boltzmann-Gibbs}
\label{ssec:level_spacing}

We now compare the introduced hyperensembles $P_{E^*}$, $P_{\gamma}$, and $P_{E^*,\alpha}$ to the Boltzmann-Gibbs distribution $\pi^\beta$. In each case we parameter-match the energetic constraint value ($E^*$ or $\overline{E}$) to the value of $E(\pi^{\beta})$, making comparisons as a function of $\beta$. Note that $P_{E^*,\alpha}$, unlike $P_{E^*}$ or $P_{\gamma}$, can recover the BG equilibrium $\pi^\beta$ in a limiting special case ($\alpha\rightarrow\infty$).

As an illustrative example, we consider a system with microstates $X=\{1,...,n\}$ and level-spaced energies: for each $x\in X$,
\begin{equation}
\label{eq:level_spaced}
\epsilon_x=\upsilon x,
\end{equation}
for a chosen spacing parameter $\upsilon>0$. The equilibrium $\pi^{\beta}$ has a truncated-geometric form,
\begin{equation}
\label{eq:pi_beta_level}
\pi^{\beta}_x=\frac{e^{-\beta \upsilon x}}{Z_\beta}, \ \ Z_\beta=\frac{1-e^{-n\beta\upsilon}}{e^{\beta\upsilon}-1},
\end{equation}
hereafter setting $\upsilon=1$ for simplicity. To compare the hyperensembles to $\pi^\beta$, we first establish the appropriate parameter-matchings.

\subsubsection{Parameter-matchings $E^*(\beta)$ and $\gamma(\beta)$}
\label{sssec:param_matchings}

For $P_{E^*}$, the parameter $E^*$ associated with a given value of $\beta$ is simply $E^*=E(\pi^\beta)$. Or, for a given $E^*$, the associated $\beta$ is obtained by solving $E^*=E(\pi^\beta)$ for $\beta(E^*)$. In the linearly spaced energies model~\eqref{eq:pi_beta_level}, we have
\begin{equation}
\label{eq:E_of_pibeta}
E(\pi^\beta)=-\frac{d}{d\beta}\log Z_\beta=\frac{1}{1-e^{-\beta}}-\frac{n}{e^{n\beta}-1},
\end{equation}
The BG distribution $\pi^{\beta(E^*)}$ can be seen as the unambiguous reference equilibrium for both $P_{E^*}$ and $P_{E^*,\alpha}$, since it is the unique point of intersection between the parametric family $(\pi^\beta)_{\beta\in\mathbb{R}}$ and the hyperensemble support $\Delta_{E^*}$. $\pi^{\beta(E^*)}$ is the mode of $P_{E^*,\alpha}$ for $\alpha>0$, and the least probable distribution for $\alpha<0$. Given $E^*$, or equivalently $\beta(E^*)$, the model $P_{E^*,\alpha}$ has only one free parameter remaining: $\alpha\in\mathbb{R}$, or equivalently $\overline{S}=\frac{\partial}{\partial\alpha}\log Z(E^*,\alpha)$. The latter has values in the range $S_{\mathrm{min}}(E^*)\le \overline{S}\le S_{\mathrm{max}}(E^*)$, with the upper bound being $S_{\mathrm{max}}(E^*)=E(\pi^\beta)$ and the lower bound given by Eq.\,\eqref{eq:min_pair}.

For the hyperensemble $P_{\gamma}$, we establish a parameter matching between $\gamma$ and $\beta$ holding throughout the range $\epsilon_-\le \overline{E}\le\epsilon_+$. We do this by first calculating $\overline{E}(\gamma)=-\frac{d}{d\gamma}\log Z(\gamma)$ after computing $Z(\gamma)$ for the case of level-spaced energies in Appendix~\ref{app:Zgamma}. Then, setting $\overline{E}(\gamma)=E(\pi^\beta)$, we obtain
\begin{equation}
\label{eq:transcendental_gamma_beta}
\frac{1}{\gamma(\beta) }-\frac{1}{e^{\gamma(\beta) }-1}=\frac{1}{n-1}\left[\frac{1}{e^{\beta }-1}-\frac{ n }{e^{n \beta }-1}\right],
\end{equation}
a transcendental with unique root $\gamma(\beta)$. Note that the $\beta=0$ limit corresponds to $\gamma=0$, with $E(\pi^0)=\overline{E}(0)=\mathcal{E}$; the latter equals $\frac{1+n}{2}$ for level-spaced energies.

\subsubsection{Nonequilibrium free energy}
\label{ssec:BG_comparison}

\begin{figure}
\includegraphics[width=0.85\linewidth,trim=10 10 10 0,clip]{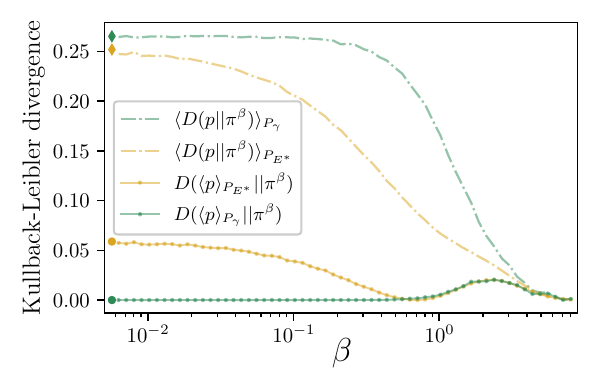}
\caption{Kullback-Leibler divergences $\langle D(p||\pi^\beta)\rangle_{P}$, and $D(\langle p\rangle_P ||\pi^\beta)$ for both energy-constrained hyperensembles, $P_{E^*}$ and $P_\gamma$, under parameter matchings of Sec.~\ref{sssec:param_matchings}. Displayed is the case of $n=3$ with $\epsilon_x=x$, with means computed from $10^5$ steps of MCMC sampling (Appendices~\ref{app:hitandrun},~\ref{app:rej_mcmc}). Dashed curves: mean KL divergence to $\pi^\beta$ for random distributions; dotted curves: KL divergence of the mean distribution to $\pi^\beta$. Associated theoretical values at $\beta=0$ (Appendix~\ref{app:beta0_KLvals}) are displayed to the left of the curves.}
\label{fig:KL}
\end{figure}

The excess nonequilibrium free energy $\Delta F_\beta(p)$ of $p\in\Delta(X)$ relative to equilibrium $\pi^\beta$ can be expressed via the Kullback-Leibler (KL) divergence, $D(p||q):=\sum_{x\in X}p_x\log\frac{p_x}{q_x}$ \cite{KL1951}. In particular, the free energy is denoted $F_\beta(p):=E(p)-\beta^{-1}S(p)$, taking equilibrium value $F_\beta(\pi^\beta)=-\beta^{-1}\log Z_\beta$; then $\Delta F_\beta(p):=F_\beta(p)-F_\beta(\pi^\beta)$ is
\begin{equation}
\begin{aligned}
\label{eq:deltaF}
    \Delta F_\beta(p)&=\sum_x p_x\left[\epsilon_x+\frac{\log Z_\beta}{\beta}\right]-\frac{S(p)}{\beta}\\
    &=\frac{1}{\beta}\sum_{x\in X}p_x\log \frac{p_x}{\pi^\beta_x}=\frac{1}{\beta}D(p||\pi^{\beta}).
\end{aligned}
\end{equation}
Thus we may sample $p$ from $P_{E^*(\beta)}$ or $P_{\gamma(\beta)}$ and examine the resulting statistics of $\Delta F_\beta(p)$; such analyses quantify how far from equilibrium one may expect a system to be, given knowledge only of the sample space, microstate energies, and exact or expected average energy. Incorporating the entropic constraint permits $\langle F_\beta(p)\rangle$ to be directly tuned; in particular,
\begin{equation}
\langle F_\beta(p)\rangle_{P_{E^*,\alpha}}=E^*-\overline{S}/\beta(E^*).
\end{equation}
Additionally, we compute the free energy of the {\it mean} distribution $\overline{p}:=\langle p\rangle_{P}$ in each hyperensemble. The effect of averaging is to reduce free energy, i.e.,
\begin{equation}
\label{eq:KL_convexity}
D(\langle p\rangle_P||\pi^\beta)\le \langle D(p||\pi^\beta)\rangle_P,
\end{equation}
following from the convexity of KL divergence~\cite{coverthomas}. The mean distribution thus can be unrepresentative from a thermodynamic standpoint, suppressing the nonequilibrium nature of random $p\sim P$. This is notable in light of past hyperstatistical modeling approaches that focused on the mean distribution as a central object of interest~\cite{Beck_2004,beck2004superstatistical,Stopnitzky2019}. The mean distribution is closer to equilibrium than typical random distributions, but nevertheless, strict inequality $\langle p\rangle_P\ne \pi^\beta$ still holds except in special cases, so even $\overline{p}$ generally has nonzero excess free energy.

We consider the mean KL divergence, and KL divergence of the mean distribution, to equilibrium, focusing on the level-spaced energies example. Appendix~\ref{app:beta0_KLvals} examines these values in the limit $\beta\rightarrow 0$, under the parameter-matchings established in Sec.~\ref{sssec:param_matchings}, including exact analytics at $n=3$; see Fig.~\ref{fig:KL}. In both models $P_{E^*}$ and $P_{\gamma}$, the quantity $\langle D(p||\pi^\beta)\rangle$ decreases monotonically from a maximum at $\beta=0$. The limiting value
\begin{equation}
\begin{aligned}
\lim_{\gamma\rightarrow 0}\langle D(p||\pi^{\beta(\gamma)})\rangle_{P_\gamma}&=\langle D(p||\pi^0)\rangle_{ \mathrm{U}[\Delta(X)]}\\
&=\log n-H_{n}+1,
\end{aligned}
\end{equation}
where $H_n=\sum_{k=1}^n\frac{1}{k}$ is the Harmonic number. The value at $n=3$ is $\log 3-\frac{5}{6}\approx 0.265$; see Fig.~\ref{fig:KL}. For $P_{E^*}$, the $\beta\rightarrow 0$ limit corresponds to $E^*\rightarrow\mathcal{E}$ (the arithmetic average energy), and
\begin{equation}
\begin{aligned}
\label{eq:lim_Estar_to_mathcal_E}
\lim_{E^*\rightarrow\mathcal{E}}\langle D(p||\pi^{\beta(E^*)})\rangle&=\langle D(p||\pi^0)\rangle_{P_{\mathcal{E}}}\\
&=\log n-\langle S(p)\rangle_{\mathrm{U}[\Delta_{\mathcal{E}}]}.
\end{aligned}
\end{equation}
At $n=3$, Eq.\,~\eqref{eq:lim_Estar_to_mathcal_E} evaluates to $\log\frac{3}{\sqrt{2}}-\frac{1}{2}\approx 0.252$. The mean distribution $\overline{p}$ has nonmonotonic KL divergence to equilibrium as a function of $\beta$ in both models. $D(\langle p\rangle_{P_{\gamma(\beta)}}||\pi^\beta)$ increases from $0$ (which follows from $\langle p\rangle_{P_0}=\pi^0$) and reaches a global maximum before decreasing back to zero as $\beta\rightarrow\infty$. The quantity $D(\langle p\rangle_{P_{E^*(\beta)}}||\pi^\beta)$ similarly has a local maximum around the same $\beta$, but also has a local minimum at smaller $\beta$. In contrast with $\langle p\rangle_{P_{\gamma(\beta)}}$, the KL divergence to equilibrium of $\langle p\rangle_{P_{E^*(\beta)}}$ remains nonzero as $\beta\rightarrow 0$ ($P_{E^*}\rightarrow P_{\mathcal{E}}$). We show 
\begin{equation}
D(\langle p\rangle_{P_{\mathcal{E}}}||\pi^0)=\log n-S(\mathrm{centroid}(\Delta_{\mathcal{E}})),
\end{equation}
which at $n=3$ equals $\log\frac{3}{2\sqrt{2}}\approx 0.058$. For further details, see Appendix~\ref{app:beta0_KLvals}.

\begin{figure}
\includegraphics[width=\linewidth,trim=5 15 0 0]{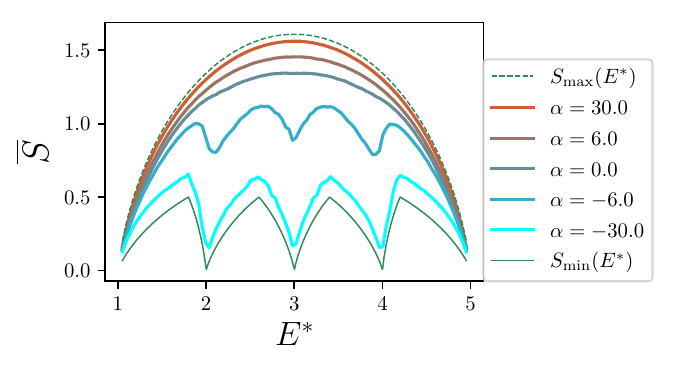}
\caption{Mean Shannon entropy $\overline{S}$ versus $E^*$ for random probability distributions from the hyperensemble $P_{E^*,\alpha}$ (Sec.~\ref{ssec:entropic_energetic}) at several values of $\alpha$, with $n=5$ and energies $\epsilon_x=x$. The value of $\overline{S}$ for $\alpha\gg 0$ is near $S_{\mathrm{max}}(E^*)$, the Boltzmann-Gibbs value; for $\alpha\ll 0$ it is near $S_{\mathrm{min}}(E^*)$ (Eq.\,\eqref{eq:S_min_linear}), varying non-monotonically with $E^*$. The mean entropy is computed from $2\times 10^5$ steps of MCMC sampling (Appendices~\ref{app:hitandrun},~\ref{app:rej_mcmc}).}
\label{fig:barS_vs_Estar_alpha}
\end{figure}

In $P_{E^*,\alpha}$, to demonstrate the parameterization of $\overline{S}$ by $\alpha$, we examine $\langle S(p)\rangle_{P_{E^*,\alpha}}$ as a function of $E^*$ at several values of $\alpha$ in Fig.~\ref{fig:barS_vs_Estar_alpha}. For $\alpha\gg 0$, the mean entropy is near that of $\pi^\beta$, namely,
\begin{equation}
S_{\mathrm{max}}(E^*)=\beta E^*+\log(1-e^{-n\beta})-\log(e^{\beta}-1),
\end{equation}
with $\beta$ the solution of $E^*=(1-e^{-\beta})^{-1}-n(e^{n\beta}-1)^{-1}$. For $\alpha<0$, at large $|\alpha|$, samples $p\sim P_{E^*,\alpha}$ are near the entropy minimizer $p_{E^*}^{\mathrm{min}}$, with $\overline{S}$ inheriting the non-monotonicity of $S_{\mathrm{min}}(E^*)$ as can be seen in Fig.~\ref{fig:barS_vs_Estar_alpha}. Making note in Eq.\,\eqref{eq:min_pair} that for $\upsilon=1$ we have $\epsilon_{<}=\lfloor E^*\rfloor$ and $\epsilon_{>}=\lceil E^*\rceil$, the minimal entropy is
\begin{equation}
\label{eq:S_min_linear}
S_{\mathrm{min}}(E^*)=\min\left(h\left(\frac{E^*-\lfloor E^*\rfloor}{n-\lfloor E^*\rfloor}\right),h\left(\frac{\lceil E^*\rceil-E^*}{\lceil E^*\rceil-1}\right)\right).
\end{equation}
Given the constraint $E(p)=E^*$, the free energy excess of any $p\sim P_{E^*,\alpha}$ relative to $\pi^{\beta}$ at $\beta=\beta(E^*)$ is given by
\begin{equation}
    \Delta F_\beta(p)=-\beta^{-1}(S(p)-S(\pi^\beta)),
\end{equation}
i.e., entirely informational. The average distribution satisfies $S(\langle p\rangle_{P_{E^*,\alpha}})\ge\overline{S}$ by concavity of entropy~\cite{coverthomas}, reflecting that $\overline{p}$ is closer to equilibrium than typical random $p$ drawn from a nonequilibrium hyperensemble.

\section{Applications}
\label{sec:applications}
Here we demonstrate applicability of the hyperstatistical modeling framework of Sec.~\ref{sec:framework} in a few areas: superstatistics (Sec.~\ref{ssec:superstatistics}), nonequilibrium steady states (Sec.~\ref{sec:applications_ness}), and Bayesian inference (Sec.~\ref{ssec:inference}).

\subsection{Superstatistics}
\label{ssec:superstatistics}

Superstatistics is a modeling framework that utilizes a Boltzmann-Gibbs distribution $\pi^\beta$ with random inverse temperature $\beta$ drawn from a prescribed density $f(\beta)$, reflecting a quasi-equilibrium setting~\cite{Leo_van_Hemmen_2013, Falcioni_2011,touchette2004temperature,Beck2011,Salazar2012,dixit2015detecting}. The mixing density $f(\beta)$ is free to choose, and can be estimated from data or derived on the basis of entropy maximization arguments~\cite{naudts2007generalised}. In a superstatistics model, the distribution $\pi^\beta$ inherits randomness from $\beta\sim f$, and thus is a hyperensemble albeit with support restricted to the one-dimensional family $\{\pi^\beta\}_{\beta\in\mathbb{R}}$. In particular,
 \begin{equation}
\label{eq:superstatistics_hyperensemble}
P_f(p)=\int_{\mathbb{R}}f(\beta)\delta(p-\pi^\beta )d\beta.
\end{equation}
However, subject to constraint $E(p)=E^*$, superstatistics can only predict equilibrium. That is because of the monotonic relationship $\beta=\beta(E^*)$, which is only compatible with $f(\beta)=\delta(\beta-\beta(E^*))$, in which case $P_f(p)=\delta(p-\pi^{\beta(E^*)})$. In contrast, the hyperensembles $P_{E^*}$ and $P_{E^*,\alpha}$ include also all nonequilibria $p\in\Delta_{E^*}\backslash\{\pi^{\beta(E^*)}\}$ in their support. Thus at fixed $E(p)=E^*$, comparison of $P_{E^*}$ or $P_{E^*,\alpha}$ to superstatistics reduces to their comparison to Boltzmann-Gibbs (Sec.~\ref{ssec:BG_comparison}).

For a more substantive comparison between superstatistics and the framework of Sec.~\ref{sec:framework}, we consider two complementary approaches. In Sec.~\ref{ssec:P_Estar_mixture_f_beta}, we take $f(\beta)$ as given and construct a mixture model of $P_{E^*(\beta)}$, akin to how superstatistics is a mixture model of $\pi^\beta$. In Appendix~\ref{app:f_beta_for_P_gamma} we examine the special case in which the density of average energy aligns with that of $P_\gamma$.

\subsubsection{Mixture model of $P_{E^*}$ for a given $f(\beta)$}
\label{ssec:P_Estar_mixture_f_beta}

In Sec.~\ref{ssec:exact_average_energy} we introduced the average energy hyperensemble $P_{E^*}$, which satisfies $E(p)=E^*$ (as does the BG equilibrium $\pi^{\beta(E^*)}$), but which lacks any constraint on entropy. Any given superstatistics model can be seen as sampling a random value of average energy $E^*\sim \mu_f$, with $\mu_f(E^*)$ determined by $f(\beta)$ (as explained below), and then imposing an entropy maximization constraint $S(p)=S_{\mathrm{max}}(E^*)$. This motivates the notion of a superstatistical mixture of $P_{E^*}$, following the same prescription of $f(\beta)$, but without Shannon entropy maximization: draw a random $\beta\sim f$, then sample $p\sim P_{E^*}$ with $E^*=E(\pi^\beta)$. In that way, we retain freedom to choose $f(\beta)$, but impose no subsequent equilibrium assumption. We denote the resulting mixture model as $P_{\mu_f}(p)$. This approach allows comparison of quasi-equilibrium models with fully nonequilibrium counterparts, subject to aligned statistics of $E(p)$. In Appendix~\ref{app:f_beta_for_P_gamma} we consider the case corresponding to $P_{\gamma}$.

The density of average energy $\mu_f:[\epsilon_-,\epsilon_+]\rightarrow\mathbb{R}_+$ associated with a given $f(\beta)$ is obtained by transformation of random variables. Namely, using $E^*=-\frac{d}{d\beta}\log Z_\beta$ at $\beta=\beta(E^*)$, we have
\begin{equation}
\label{eq:mu_Estar_from_f_beta}
\mu_f(E^*)=\frac{f(\beta(E^*))}{V(\pi^{\beta(E^*)})},
\end{equation}
with $V(\pi^\beta)$ the energy variance at equilibrium,
\begin{equation}
\label{eq:energy_variance}
V(\pi^\beta):=\sum_{x\in X}\epsilon_x^2\pi^\beta_x-\left(\sum_{x\in X}\epsilon_x\pi^\beta_x\right)^2=\frac{d^2 \log Z_\beta}{d\beta^2}\ge 0.
\end{equation}
The resulting hyperensemble has support over all distributions with average energy values in the support of $\mu_f$. Since $P_{E^*}$ has support only on $\Delta_{E^*}$, and $E^*$ has a monotonic relationship with $\beta$, the mixture model can be represented directly as
\begin{equation}
\begin{aligned}
P_{\mu_f}(p)&=\int_{\epsilon_-}^{\epsilon_+} \mu_f(E^*)\frac{1_{E^*}(p)}{|\Delta_{E^*}|}dE^*\\
&=\frac{f(\beta(E(p)))}{|\Delta_{E(p)}| V(\pi^{\beta(E(p))})}.
\end{aligned}
\end{equation}
Sampling of the hyperensemble $P_{\mu_f}$ is a simple three-step procedure: (1) draw $\beta\sim f$, (2) calculate $E^*=E(\pi^\beta)$, then (3) sample $p\sim P_{E^*}$ via $\mathbf{HR}(\Delta_{E^*})$. The mean distribution $\overline{p}$ is again a natural object of interest; for $P_{\mu_f}$, it is a mixture of the centroids of $\Delta_{E^*}$ across $E^*$, weighted by $\mu_f(E^*)$.

\begin{figure}
\includegraphics[width=\linewidth,trim=300 50 300 50,clip]{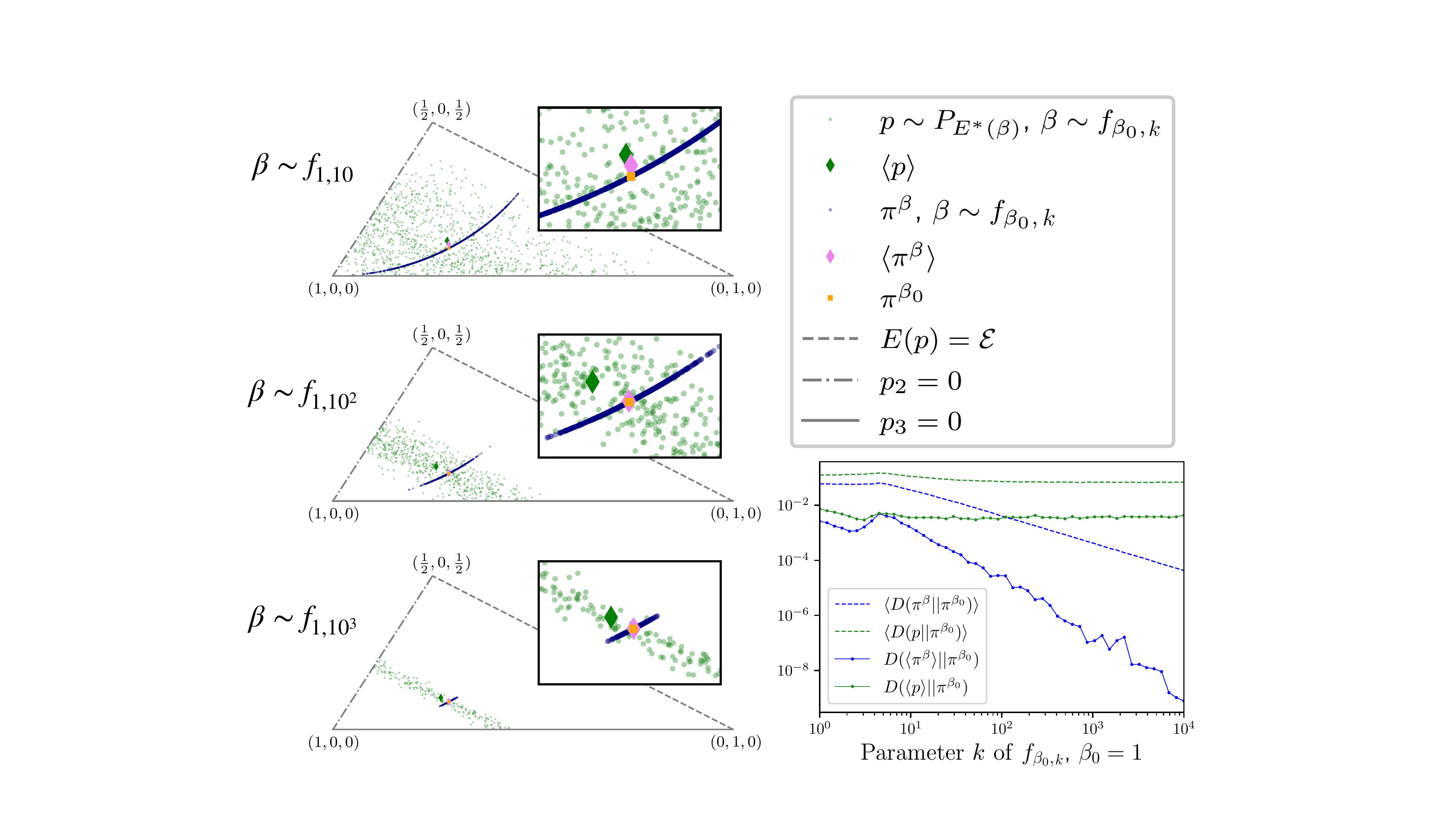}
\caption{Random samples and the mean distribution from the $f_{\beta_0,k}$-superstatistics model and the associated mixed-$P_{E^*}$ hyperensemble (Sec.~\ref{ssec:P_Estar_mixture_f_beta}), for the case of $n=3$ with energies $\epsilon_x=x$. The left three visualizations are at $\beta_0=1$ and $k=10$ (upper) $k=10^2$ (middle) and $k=10^3$ (lower). The region of the unit simplex displayed corresponds to $\beta\ge 0$ or equivalently $E^*\le \mathcal{E}=2$. Insets display the region around $\pi^{\beta_0}$. Lower right: KL divergence, on average, and of the mean distribution, to $\pi^{\beta_0}$, in both models. For superstatistics, both converge to zero as a function of the concentration parameter $k\rightarrow\infty$. For the mixture hyperensemble, both converge to their values in $P_{E^*}$ with $E^*=E(\pi^{\beta_0})$ (see Sec.~\ref{ssec:BG_comparison}). }
\label{fig:sup}
 \end{figure}

We consider the example of a $\chi^2$ density of $\beta$, under the parameterization $f(\beta)=f_{\beta_0,k}(\beta)$ from Ref.~\cite{Beck_2004}, in terms of the mean $\beta_0:=\int_0^\infty f(\beta)\beta d\beta$ and a parameter $k\in\mathbb{N}$ that controls the concentration (the variance is $2\beta_0^2/k$). Namely,
\begin{equation}
\label{eq:f_beta0_n}
f_{\beta_0,k}(\beta)=\frac{\left(\frac{k}{2}\right)^{\frac{k}{2}}}{\Gamma\left(\frac{k}{2}\right)\beta_0}\left(\frac{\beta}{\beta_0}\right)^{\frac{k}{2}-1}e^{-\frac{k}{2}\frac{\beta}{\beta_0}}, \quad \beta\ge 0.
\end{equation}
See Fig.~\ref{fig:sup} for a comparison of superstatistics and the mixed-$P_{E^*}$ model across parameter $k$ at fixed $\beta_0$.

\subsection{Nonequilibrium steady states}
\label{sec:applications_ness}

The hyperensembles described above provide static probabilistic models of nonequilibrium. They represent uncertainty over a one-time distribution $p \in \Delta(X)$ without imposition of maximum Shannon entropy. A natural next question is dynamical: how should distributional uncertainty be applied to dynamical processes? Here, we describe the corresponding construction for finite-state stochastic dynamics, focusing on nonequilibrium steady states (NESS) in discrete time Markov chains (DTMCs). This is in contrast with the common use of continuous time in stochastic thermodynamics~\cite{Seifert_2012}, which enable more realistic models and constraints~\cite{Wolpert_2019}. Yet, the following considerations are applicable to the jump chain of a stationary continuous-time Markov chain, or to the joint distribution of microstates at several instants of time. Several variational techniques have been considered in nonequilibrium modeling, such as maximum caliber~\cite{Jaynes1985, dixit2010inferring, ghosh2020} and maximum entropy production~\cite{martyushev2006,dyke2010,Endres_2017}; our approach is via $\mathbb{S}$-maximization as per Sec.~\ref{sec:framework}.

We consider stationary DTMCs over a finite sample space $X$ with transition matrix $T=(T_{xx'})_{(x,x')\in X^2}$ and stationary distribution $\rho=(\rho_x)_{x\in X}$, related by $\rho_x = \sum_{x'\in X}\rho_{x'} T_{x'x}$. We assume that the reverse of every transition is also allowed, i.e., $T_{xx'}>0\Leftrightarrow T_{x'x}>0$. The natural dynamical analogue to a microstate distribution $p$ is not just the stationary distribution $\rho$, but the two-timestep transition distribution $m$:
\begin{equation}
\label{eq:mxxprime}
    m_{xx'} := \rho_xT_{xx'}=\mathbb{P}(x_t=x,x_{t+1}=x'),
\end{equation}
i.e., the two-time marginal: the stationary probability
that the process undergoes transition $x\to x'$. Stationarity is encoded by the condition $\sum_{x\in X}(m_{xx'}-m_{x'x})=0$ holding for each $x'\in X$. A nonequilibrium steady state (NESS) is not simply a nonequilibrium distribution of microstates with respect to a given Hamiltonian; it is a stationary circulation pattern, as captured by the natural nonequilibrium observable $\Sigma(m)$, the steady-state entropy production rate:
\begin{equation}
\label{eq:sigma_m}
\Sigma(m)=\sum_{(x,x')\in X^2}m_{xx'}\log\frac{m_{xx'}}{m_{x'x}}\ge 0.
\end{equation}
NESS is characterized by $\Sigma(m)>0$; for $\Sigma(m)=0$, the dynamics is reversible, i.e., satisfies detailed balance: $m_{xx'}=m_{x'x}$ for all $(x,x')\in X^2$.

\subsubsection{Random NESS}

We apply the framework of Sec.~\ref{sec:framework} to sample random transition distributions $m$ subject to constraints. We begin by examining the set of all stationary DTMCs, demonstrating probability $1$ of NESS. Any transition distribution $m$ resides in the probability simplex $\Delta(X\times X)$; stationarity implies furthermore that $m$ resides in the intersection of $n:=|X|$ hyperplanes defining the stationary-flow polytope by
\begin{equation}
\begin{aligned}
    \mathcal{M}_{\rm sta}
    :=
    \Bigl\{\, m \;:\;&
    m\in [0,1]^{n\times n},\\
    &
    \sum_{(x,x')\in X^2}m_{xx'}=1,\\
    &
    \sum_{x'\in X}(m_{xx'}-m_{x'x})=0
    \quad \forall x\in X
    \Bigr\}.
\end{aligned}
\end{equation}

If only some transitions are allowed, as described by a set of directed edges in a transition graph with $n$ vertices, the polytope $\mathcal{M}_{\rm sta}$ is restricted to the corresponding set of directed edges. Models of random transition distributions $m\in\mathcal{M}_{\rm sta}$ can be formulated according to the framework of Sec.~\ref{sec:framework}; in the absence of any constraints beyond stationarity, the $\mathbb{S}$-maximizer is uniform on $\mathcal{M}_{\rm sta}$:
\begin{equation}
\label{eq:P_sta_uniform}
P_{\rm sta}(m)=\frac{1}{|\mathcal{M}_{\rm sta}|},
\qquad m\in\mathcal{M}_{\rm sta}.
\end{equation}
By imposing additional exact constraints of the form $Q(m)=Q^*$, $m$ is restricted to
\begin{equation}
\label{eq:M_Qstar}
\mathcal{M}_{Q^*}:=\{m\in\mathcal{M}_{\rm sta}:Q(m)=Q^*\},
\end{equation}
on which the $\mathbb{S}$-maximizing hyperensemble is again uniform. Moreover, we may impose expectation-constraints on $m$ of the form $\langle F(m)\rangle=\overline{F}$ as in Eq.\,\eqref{eq:Fj} to obtain exponential-form densities of $m$, with Lagrange multipliers determined by~\eqref{eq:dlogZdlambda}.

Detailed balance is a stronger condition than stationarity, so $\mathcal{M}_{\rm rev}\subseteq\mathcal{M}_{\rm sta}$, where the reversible-flow polytope is defined by
\begin{equation}
\mathcal{M}_{\rm rev}:=\{m\in\mathcal{M}_{\rm sta}:m_{xx'}=m_{x'x}\ \forall \ x,x'\in X\}.
\end{equation}
Moreover, transition distributions sampled from $P_{\mathrm{sta}}$ are NESS with probability $1$. We demonstrate that $\mathcal{M}_{\rm rev}$ occupies a vanishing volume fraction of $\mathcal{M}_{\rm sta}$ by counting the independent degrees of freedom of distributions in each. We assume that the transition graph is connected and that for every non-self edge present, the reverse edge is also present. Let $e$ be the number of undirected non-self edges and let $\ell$ be the number of self-loops; then $m$ then has $2e+\ell$ entries.

Of the $n$ stationarity conditions, one follows from the others, ensuring that the total probability is conserved. Together with normalization, this gives
\begin{equation}
\label{eq:dim_M}
\mathrm{dim}(\mathcal{M}_{\rm sta})=2e+\ell-n.
\end{equation}
For reversible flows, the two directions of each non-self edge have the same value. There are therefore $e+\ell$ entries before normalization, giving
\begin{equation}
\mathrm{dim}(\mathcal{M}_{\rm rev})=e+\ell-1.
\end{equation}
The difference between the two dimensionalities is
\begin{equation}
\mathrm{dim}(\mathcal{M}_{\rm sta})-\mathrm{dim}(\mathcal{M}_{\rm rev})=e-n+1,
\end{equation}
which is the cycle rank of the underlying undirected graph after the self-loops are removed, which counts the number of independent cycles~\cite{Harary1969,Diestel}. Thus the additional stationary degrees of freedom are the independent cycle currents. When the graph is a tree, stationarity implies reversibility. For the complete graph, $\mathrm{dim}(\mathcal{M}_{\rm sta})=n^2-n$, while $\mathrm{dim}(\mathcal{M}_{\rm rev})=\binom{n+1}{2}-1$, with difference $\binom{n-1}{2}$. Removal of self-loops lowers dimensionality of both $\mathcal{M}_{\mathrm{sta}}$ and $\mathcal{M}_{\mathrm{rev}}$ equivalently, leaving their difference unchanged.

\subsubsection{Three-state system}
\label{ssec:ness_three_state}

We now take $X=\{1,2,3\}$, allow all non-self transitions, and set
$m_{xx}=0$. In this case $\mathrm{dim}(\mathcal{M}_{\rm sta})=3$ and $\mathrm{dim}(\mathcal{M}_{\rm rev})=2$. We choose the cycle orientation $1\to2\to3\to1$.
Stationarity requires the three oriented edge currents to be equal:
\begin{equation}
\label{eq:three_state_current}
J:=m_{12}-m_{21}=m_{23}-m_{32}=m_{31}-m_{13}.
\end{equation}
We also define the total flows along the three pairs of opposite edges
by $a:=m_{12}+m_{21}$, $b:=m_{23}+m_{32}$, and
$c:=m_{31}+m_{13}$. Every $m\in\mathcal{M}_{\rm sta}$ can then be
written as
\begin{equation}
\label{eq:three_state_flow_parameterization}
m(a,b,c,J)=
\frac{1}{2}
\begin{pmatrix}
0&a+J&c-J\\
a-J&0&b+J\\
c+J&b-J&0
\end{pmatrix}.
\end{equation}
The variables satisfy $a+b+c=1$ and $a,b,c\geq |J|$. It follows that $|J|\leq 1/3$. 

Consider $m$ drawn from the uniform stationary-flow hyperensemble $P_{\mathrm{sta}}$. In the event of $J=J^*$, the allowed values of $(a,b,c)$ form a triangle; at $J=\pm 1/3$, it reduces to a single point corresponding to one of the two deterministic directed cycles; at $J=0$, this is the reversible triangle
$\mathcal{M}_{\rm rev}$. The triangle's side lengths are proportional to $1-3|J^*|$, and its area is therefore proportional to $(1-3|J^*|)^2$. Normalizing over $J^*\in[-1/3,1/3]$ gives the probability density of current $J$ for $m\sim P_{\mathrm{sta}}$:
\begin{equation}
\label{eq:three_state_current_density}
\nu_J(J^*)=\frac{9}{2}(1-3|J^*|)^2, \quad J^*\in\left[-\frac{1}{3},\frac{1}{3}\right].
\end{equation}
The reversible point $J=0$ is the mode, but has probability zero.

We next suppose that the stationary distribution is known exactly: $\rho(m)=\rho^*:=(\frac{1}{3},\frac{1}{3},\frac{1}{3})$. The flows satisfying this constraint form $\mathcal{M}_{\rho^*}:=\{m\in\mathcal{M}_{\rm sta}: \rho(m)=\rho^*\}$. The three outgoing marginals are $(a+c)/2$, $(a+b)/2$, and $(b+c)/2$. The constraint therefore fixes $a=b=c=1/3$, leaving only the cycle current $J$. Thus every allowed flow has the form
\begin{equation}
\label{eq:uniform_rho_flow}
m(J)=
\frac{1}{6}
\begin{pmatrix}
0&1+3J&1-3J\\
1-3J&0&1+3J\\
1+3J&1-3J&0
\end{pmatrix},
\end{equation}
so $\mathcal{M}_{\rho^*}$ is a line segment on which $\mathcal{M}_{\rho^*}$ is  uniform; equal intervals of $J$ give equal lengths along the segment, and hence
\begin{equation}
\label{eq:uniform_rho_current_density}
J\sim\mathrm{U}\left[-\frac{1}{3},\frac{1}{3}\right].
\end{equation}
As such, the steady current remains uncertain even with a constrained stationary distribution $\rho=\rho^*$; indeed, the current density changed from Eq.\,\eqref{eq:three_state_current_density} to the uniform~\eqref{eq:uniform_rho_current_density}, with the latter having $\langle |J|\rangle=1/6$ in contrast with the former's $\langle |J|\rangle=1/12$.

This example demonstrates that uncertainty over stationary dynamics remains even when the one-time distribution is known exactly, and moreover that by constraining the stationary distribution, larger expected currents can result. More generally, exact or expectation constraints on stationary distributions, cycle currents, individual edge flows, or entropy production define hyperensembles on $\mathcal{M}_{\rm sta}$. This approach can therefore provide reference models for nonequilibrium steady states in a wide range of more general contexts.

\subsection{Bayesian inference}
\label{ssec:inference}

There are many natural inference scenarios for the framework of Sec.~\ref{sec:framework}; in the simplest case, a hyperensemble $P(p)$ can be taken as a Bayesian prior, representing entirely epistemic uncertainty of a single ground-truth distribution. (In contrast, in many examples of interest, a full hyperensemble $P(p)$ may represent the ground truth, capturing population-level statistics of distributions; see Sec.\,\ref{ssec:scope}.) In this section, we consider standard Bayesian inference under observation of $N$ i.i.d.\ samples from an individual distribution; the resulting posterior hyperensemble tilts towards the observed frequencies, concentrating on a single point as $N\rightarrow\infty$. 

For a given hyperensemble $P(p)$ and an observed sequence of microstates $\mathbf{x}=\{x_i\}_{i=1}^N$, the hyperensemble posterior $P(p|\mathbf{x})$ as a function of $\mathbf{x}$ is given by Bayes's rule. Random samples from $P(p|\mathbf{x})$ may be drawn via MCMC (Appendix~\ref{app:rej_mcmc}), and its mode $p^*$ and mean $\overline{p}$ can be characterized analytically. The results below are for {\it fixed} $\mathbf{x}$, but we may also invoke {\it random} $\mathbf{x}$, e.g., to show that when $\mathbf{x}$ are drawn i.i.d.\ from a fixed $q$ in the support of $P$, perfect inference is possible as $N\rightarrow\infty$, washing out the prior entirely. For finite $N$, generating random $\mathbf{x}$ allows examination of generic behavior. Note, then, that even the mean and mode of $P(\cdot|\mathbf{x})$ become random, inheriting their randomness from $\mathbf{x}$.

The i.i.d.\ measurement model allowing a Bayesian posterior is as follows: given a categorical distribution $p$, the probability of $\mathbf{x}$ is
\begin{equation}
p_{\mathbf{x}}=\prod_{x\in X}p_x^{N_x},
\end{equation}
with $N_x:=\sum_{i=1}^N\mathds{1}\{x_i=x\}$ denoting the empirical count of $x$ in $\mathbf{x}$. The posterior hyperensemble $P(p|\mathbf{x})$ is then
\begin{equation}
P(p|\mathbf{x})=\frac{P(p)p_{\mathbf{x}}}{\langle p_{\mathbf{x}}\rangle},
\end{equation}
where the mean probability of $\mathbf{x}$ under the prior is is
\begin{equation}
\langle p_{\mathbf{x}}\rangle:=\int_{\Delta(X)}P(p)p_{\mathbf{x}}dp=\int_{\Delta(X)}P(p)\prod_{x\in X}p_x^{N_x}dp.
\end{equation}
We consider $P(p)$ subject to expectation constraints $\langle F(p)\rangle=\overline{F}$ with Lagrange multipliers $\lambda$, yielding $P_\lambda(p)=e^{\lambda\cdot F(p)}/Z(\lambda)$ as per Sec.~\ref{sec:framework}. The posterior hyperensemble then has the form
\begin{equation}
\label{eq:hyp_posterior}
P_\lambda(p|\mathbf{x})=\frac{e^{\lambda \cdot F(p)}\prod_{x\in X}p_x^{N_x}}{Z(\lambda,\mathbf{x})},
\end{equation}
with
\begin{equation}
\label{eq:Z_lam_x}
Z(\lambda,\mathbf{x})=\int_{\Delta(X)}e^{\lambda \cdot F(p)+\sum_{x\in X}N_x \log p_x}dp.
\end{equation}
The posterior can be sampled via MCMC~\cite{tierney1994markov}. The mode of $P(p|\mathbf{x})$, i.e., the maximum a posteriori estimate of $p$~\cite{dashti2013map}, is
\begin{equation}
\label{eq:posterior_mode}
p^*=\underset{p\in \Delta(X)}{\mathrm{argmax}}\, \left(\lambda \cdot F(p)+\sum_{x\in X}N_x \log p_x\right).
\end{equation}
The posterior mean $\overline{p}=\int_{\Delta(X)}pP(p|\mathbf{x})dp$~\cite{ericson1970posterior} is expressed elementwise by
\begin{equation}
\label{eq:posterior_mean}
\overline{p}_{x'}=\frac{Z(\lambda,\mathbf{x}\sqcup x')}{Z(\lambda,\mathbf{x})},
\end{equation}
with $\mathbf{x}\sqcup x':=\{x_1,....,x_N,x'\}$ denoting a vector of microstates of length $N+1$ consisting of $\mathbf{x}$ with an appended $(N+1)$st element, $x_{N+1}=x'$. For large $N$, field-theoretic techniques can be used to estimate $\overline{p}$~\cite{Bialek1996}.

\subsubsection{Inference under Crooks's prior} 
\label{ssec:inference_crooks}
Crooks's hyperensemble~\cite{crooks2007beyond} (Sec.~\ref{sec:entropic_constraint}, Appendix~\ref{app:crooks}) takes the form $P_{\gamma,\alpha}(p)=e^{-\gamma F_{\gamma/\alpha}(p)}/Z(\alpha,\gamma)$, with $(\gamma,\alpha)$ determining the values of constrained expectations $\langle E(p)\rangle=\overline{E}$ and $\langle S(p)\rangle=\overline{S}$; for $\beta=\alpha/\gamma>0$, the mode of $P_{\gamma,\alpha}$ is $\pi^{\beta}$. Treating $\pi^{\beta}$ as a reference equilibrium, predictions from $P_{\gamma,\alpha}$ are generically nonequilibrium, with mean free energy
\begin{equation}
\langle \Delta F_\beta\rangle_{P_{\gamma,\alpha}}=\overline{E}-\beta^{-1}\overline{S}.
\end{equation}
We augment Crooks's hyperensemble by incorporation of observed data $\mathbf{x}=\{x_i\}_{i=1}^N$; the hyperensemble posterior~\eqref{eq:hyp_posterior} is then
\begin{equation}
\begin{aligned}
\label{eq:posterior_crooks}
P_{\gamma,\alpha}(p|\mathbf{x})&=\frac{e^{-\gamma E(p)+\alpha S(p)}\prod_{x\in X}p^{N_x}}{Z(\gamma,\alpha,\mathbf{x})}\\
&=\frac{e^{\sum_{x\in X}\left[ -\gamma p_x\epsilon_x- \alpha p_x\log p_x+ N_x\log p_x\right]}}{Z(\gamma,\alpha,\mathbf{x})},
\end{aligned}
\end{equation}
which can be sampled from straightforwardly by rejection and MCMC (Appendix~\ref{app:rej_mcmc}). Hence properties of $P(p|\mathbf{x})$ can be estimated to high precision, e.g., the posterior mean distribution $\overline{p}$ (Eq.~\eqref{eq:posterior_mean}) or the mean free energy excess $\beta^{-1}\langle D(p||\pi^{\beta})\rangle$. The posterior mode $p^*$ (Eq.~\eqref{eq:posterior_mode}) is obtained by maximizing the exponent in the second expression of~\eqref{eq:posterior_crooks}. Introducing a Lagrange multipler $\psi$ for normalization, the stationarity condition with respect to $p_x$ is
\begin{equation}
\label{eq:mode_stationary}
-\gamma\epsilon_x-\alpha(1+\log p_x)+\frac{N_x}{p_x}+\psi=0.
\end{equation}
Denoting $z=e^{1-\frac{\psi}{\alpha}}$, the modal probability $p^*_x$ solving~\eqref{eq:mode_stationary} is, if $N_x>0$,
\begin{equation}
p^*_x=\frac{N_x}{\alpha W(\frac{z}{\alpha}e^{\beta\epsilon_x}N_x)},
\end{equation}
with $W$ denoting the Lambert $W$ function's principal branch; if $N_x=0$, the condition~\eqref{eq:mode_stationary} simplifies and $p^*_x$ is
\begin{equation}
p^*_x=\frac{e^{-\beta\epsilon_x}}{z},
\end{equation}
i.e., proportional to $\pi^{\beta}_x$ across microstates not appearing in $\mathbf{x}$. The value of $z$ is uniquely determined by the normalization condition, $\sum_{x\in X}p^*_x=1$. At $N=0$, the original mode is recovered ($p^*=\pi^{\beta}$) and $z=Z_{\gamma/\alpha}$. Note that $p^*$ can be expressed in BG form as $\tilde{\pi}^{\beta}:=\frac{1}{z}e^{-\beta\tilde{\epsilon}_x}$ with $\tilde{\epsilon}_x=\epsilon_x-\beta^{-1}W(\frac{z}{\alpha}e^{\beta\epsilon_x}N_x)$. In Appendix~\ref{app:mode_F} we show that the KL divergence between the reference equilibrium $\pi^{\beta}$ and the posterior mode $p^*$ is given by
\begin{equation}
D(p^*||\pi^{\beta})=\frac{N}{\alpha}+\log\frac{Z_{\beta}}{z},
\end{equation}
or equivalently, that $F_{\beta}(p^*)=\frac{1}{\gamma}(N-\alpha\log z)$.

\section{Discussion}
\label{sec:discussion}

In this work, we have formulated models of random probability distributions $P(p)$ subject to constraints on exact or expected values of properties of $p$. A hyperensemble $P(p)$ maximizing differential entropy $\mathbb{S}(P)$ subject to incomplete knowledge provides a reference model for distributional statistics and a prior for statistical inference. This approach follows the conventional modeling logic of Jaynes and is applicable whenever the sample space itself is a probability simplex $\Delta(X)$. Distributional uncertainty is already familiar across domains, most plainly in Bayesian statistics \cite{van_de_Schoot_2021}; indeed, each considered hyperensemble is a categorical prior. Random distributions also arise in, e.g., models of disordered systems in physics~\cite{Bovier_2006}, empirical sampling distributions in large deviation theory~\cite{den_Hollander_2008}, and random measures in probability theory~\cite{Kallenberg_2017}. The framework of Sec.~\ref{sec:framework} generalizes the entropic prior and Crooks's hyperensemble; we examined $P_{E^*}$, $P_{\gamma}$, and $P_{E^*,\alpha}$, comparing them to Boltzmann-Gibbs (Sec.~\ref{sec:therm_hyp}) and demonstrating several applications (Sec.~\ref{sec:applications}). Numerous theoretical extensions would be natural, e.g., exploring other constraint choices for random NESS (Sec.~\ref{sec:applications_ness}), or furthering the geometric characterization of $\Delta_{E^*}$ (Sec.~\ref{ssec:exact_average_energy})~\cite{webb1996central,lasserre2015volume,Dirksen2017}.

In Sec.~\ref{ssec:inference} we considered the simplest inference scenario for this framework, but one may complicate it by including noise or correlations into the measurement model, or considering inference subject to exact constraints such as $E(p)=E^*$~\cite{Silva_2015,polson2025bayesian} or relaxed forms thereof~\cite{Duan2020,Presman2023}. The parameters of hyperensemble $P(p)$ may also be uncertain; we consider uncertain $E^*$ in $P_{E^*}$ (Sec.~\ref{ssec:P_Estar_mixture_f_beta}), and uncertainty in $\alpha$ of the entropic prior $P_\alpha$ (equivalently, an uncertain $\overline{S}$) has been considered~\cite{Skilling1989,Strauss1993}. Indeed, the formulation of a random $P$ may be even approached by application of modeling logic analogous to that of Sec.~\ref{sec:framework}. However, the set of possible $P$ is uncountably infinite-dimensional, so to assign a prior one must first specify a parametric family. One may continue to introduce hierarchical Bayesian layers indefinitely, but as we have demonstrated in this work, even a single layer of distributional uncertainty allows access to meaningful constraints on $S(p)$, $E(p)$, and other nontrivial properties of $p\sim P$.

For models in statistical mechanics, the choice of $S(p)=S_{\mathrm{max}}$ can be seen as a {\it physical} assumption; removing or relaxing it permits straightforward modeling of nonequilibrium statistics, at the mere cost of no longer suppressing distributional randomness. We emphasize this work's compatibility with established approaches in nonequilibrium physical modeling~\cite{ruelle2003extending,ryazanov2009lifetime,otten2010,Habeck2014,truffet2018shannon,Shen2017,Tsallis2019}, with many intersections explorable in future works. For instance, one may formulate models of random density matrices~\cite{wootters1990random,zyczkowski2001induced,breuer2007} via this approach; some constrained sampling algorithms are already known~\cite{Zyczkowski2011generating,Alsing2022}. As another example, we have considered constraints on Shannon entropy $S(p)$, but the approach is equally applicable to other entropies, e.g., R\'enyi or Tsallis~\cite{renyi1961measures,Tsallis_1988}. This could allow connections with nonextensive statistical mechanics~\cite{Jizba2019,korbel2018classification,Tsallis2019}, and moreover, mixed hyperensembles thereof (a la Sec.~\ref{ssec:superstatistics}) relate to the Tsallis extension of superstatistics~\cite{squillante2026hyperstatistics}. (Similarly, $\mathbb{S}$ could be modified in the variational principle of Sec.~\ref{sec:framework}.) As one more example from nonequilibrium physics, consider the setting of information thermodynamics~\cite{Parrondo_2015,Strasberg2017}; observe that noisy-channel measurements induce a {\it random} posterior distribution, as implied by Bayes's rule and the random measurement outcome. The posterior has larger free energy by $\beta^{-1}I$ {\it on average}, with $I$ denoting the mutual information of the measurement channel, but the full statistics of free energy can be examined in reference to chosen nonequilibrium hyperensembles. More generally, we anticipate similar intersections and applicability throughout physics, data science, statistics, and probabilistic modeling writ large.

Our approach provides mechanism-free predictions via the logic of null modeling; whenever imposed constraints leave open a degeneracy of distributions, one may formulate priors according to this framework. Yet, one may ask if $P(p)$ and the constraints defining it are entirely epistemic devices, or if they have meaningful physical interpretations. As discussed in Sec.~\ref{ssec:scope}, numerous datasets consist of heterogeneous populations of distributions, and in such cases the formulated $P(p)$ represent a model of a physically present `ensemble of ensembles'. In the energetically constrained hyperensembles of Sec.~\ref{sec:therm_hyp}, a notion of `energy balance' is present, but without driving the system towards equilibrium (instead, $\Delta_{E^*}$ is traversed). The entropic constraint $\langle S(p)\rangle=\overline{S}$ for $\overline{S}<S_{\mathrm{max}}$ can be seen as indicating a `source of nonequilibrium' (e.g., from random driving~\cite{Verley_2014} or disequilibrium conversion~\cite{BRANSCOMB20171}), that is, in relation to a default assumption of $\overline{S}=S_{\mathrm{max}}$. But from the generalized standpoint of distributional statistics, no such `source' need be assumed; the entropy constraint can be removed entirely, and nonequilibrium is expected by default. In the absence of complete information or an equilibrium assumption, distributional randomness and the associated nonequilibrium statistics naturally arise.

\renewcommand{\addcontentsline}[3]{}

\bibliography{bib}

\appendix

\section{Maximization of differential entropy}
\label{app:entropy}

Here we consider the variational problem of formulating a probability density of distributions subject to exact and expectation constraints as per Sec.~\ref{sec:framework}. A priori we take background measure $M(p)$ whose support is the unit simplex $\Delta(X)$ of categorical distributions over $X$, with $n:=|X|$ finite. $M(p)$ is normalized as
\begin{equation}
1=\int_0^1dp_1\int_0^{1-p_1}dp_2\cdots \int_0^{1-\sum_{i=1}^{n-2}p_i}dp_{n-1}\, M(p),
\end{equation}
where the argument $p$ of $M(p)$ is $p=(p_1,...,p_n)$ with $p_n:=1-\sum_{i=1}^{n-1}p_i$. We maximize the differential entropy, or more specifically, the negative KL divergence with respect to the background measure:
\begin{equation}
\mathbb{S}(P)=-\int_{\Delta(X)}P(p)\log\frac{P(p)}{M(p)}dp,
\end{equation}
subject to constraints $Q(p)=Q^*$ (which restrict the support from $\Delta(X)$ to $\Delta_{Q^*}$). $\Delta_{Q^*}$ may be of lower dimensionality, in which case $dp$ is the associated surface element. Additionally, $P$ is subject to expectation constraints of the form
\begin{equation}
\langle F(p)\rangle:=\int_{\Delta_{Q^*}}P(p)F(p)dp=\overline{F},
\end{equation}
where $F(p)$ is a real-valued vector function of $p$; in addition, we require normalization,
\begin{equation}
\langle 1\rangle=\int_{\Delta_{Q^*}}P(p)dp=1.
\end{equation}
This yields a Lagrangian for maximization, 
\begin{equation}
L(P)=\mathbb{S}(P)+\lambda\cdot( \langle F(p)\rangle-\overline{F})+\zeta (\langle 1\rangle-1).
\end{equation}
We consider the function $P(p)$ as a constrained collection of values $\{P(p)\}_{p\in\Delta_{Q^*}}$, and maximize $L(P)$ with respect to each given value $P(p)$, as
\begin{equation}
\frac{\partial L(P)}{\partial P(p)}=-\log\frac{P(p)}{M(p)}-1+\lambda \cdot F(p)+\zeta=0.
\end{equation}
Denoting $Z=e^{1-\zeta}$, the unique $\mathbb{S}$-maximizer is
\begin{equation}
P(p)=Z^{-1}M(p)e^{\lambda \cdot F(p)}1_{Q^*}(p),
\end{equation}
where
\begin{equation}
Z(Q^*,\lambda) := \int_{\Delta_{Q^*}}M(p)e^{\lambda \cdot F(p)}dp
\end{equation}
and $1_{Q^*}(p):=\mathds{1}\{Q(p)=Q^*\}$. The Lagrange multipliers are determined by $\nabla_\lambda \log Z = \overline{F}$, and the support of $P$ is restricted to $\Delta_{Q^*}$ by the factor $1_{Q^*}(p)$.

We herein consider the Euclidean metric yielding background measure $M(p)=\mathrm{const}$, so that we may write
\begin{equation}
\begin{aligned}
P(p) &= Z(Q^*,\lambda)^{-1}e^{\lambda \cdot F(p)}1_{Q^*}(p), \\
Z(Q^*,\lambda) &= \int_{\Delta_{Q^*}}e^{\lambda \cdot F(p)}dp.
\end{aligned}
\end{equation}

\section{$\mathrm{Dirichlet}(\mathbf{1})$, entropic prior, and Crooks's hyperensemble}
\label{app:dirichlet_entropic_crooks}

Herein we briefly provide background on $\mathrm{Dirichlet}(\mathbf{1})$ (the uniform $\mathrm{U}[\Delta(X)]$), $P_{\alpha}$ (the entropic prior) and $P_{\gamma,\alpha}$ (Crooks's hyperensemble).

\subsection{$\mathrm{Dirichlet}(\mathbf{1})$}
\label{app:dirichlet}

The simplest hyperensemble is the Euclidean uniform prior over the simplex $\mathrm{U}[\Delta(X)]$, i.e., the $\mathrm{Dirichlet}(\mathbf{1})$ distribution (for a general Dirichlet distribution with parameters $\{\alpha_x\}_{x\in X}$, this corresponds to $\alpha_x=\mathrm{const}=1$ w.l.o.g. when selecting value $1$; we use notation $\mathbf{1}$ for the vector of ones). The associated hyperensemble is uniform with density for all $p\in \Delta(X)$,
\begin{equation}
P(p)=(n-1)!,
\end{equation}
where $n:=|X|$. This model can be sampled by drawing i.i.d.\ exponentials and rescaling them to unit sum. This model has $\langle E(p)\rangle=\mathcal{E}:=\frac{1}{n}\sum_{x\in X}\epsilon_x$, and $\langle S(p)\rangle=H_{n}-1$, approaching $1-\gamma_E$ as $n\rightarrow\infty$, where $\gamma_E\approx 0.577$ is the Euler-Mascheroni constant.

\subsection{Entropic prior}
 \label{app:entropic}
 
 We may equally well consider {\it expectation}-based entropic constraints, with knowledge encoded as expected Shannon entropy
\begin{equation}
\langle S(p)\rangle=\overline{S}\in[0,\log n],
\end{equation}
leading to an exponential-family hyperensemble with entropy-dependent probability density:
\begin{equation}
\label{eq:entropic_prior}
P_\alpha(p)=\frac{e^{\alpha S(p)}}{Z(\alpha)}, \ \ \ \ Z(\alpha)=\int_{\Delta(X)}e^{\alpha S(p)}dp,
\end{equation}
in statistics called the {\it entropic prior} \cite{Neumann2007,Brand1999} for a categorical distribution. The extrema $\overline{S}=0$ and $\overline{S}=\log n$ correspond to $\alpha=-\infty$ and $\alpha=\infty$, respectively, and $\alpha=0$ corresponds to $\mathrm{Dirichlet}(\mathbf{1})$. $P_\alpha$ can be sampled from exactly by an elementary rejection algorithm (Appendix~\ref{app:rej_mcmc}).

\subsection{Crooks's hyperensemble}
\label{app:crooks}

Crooks introduced a hyperensemble in which both $S(p)$ and $E(p)$ are subject to expectation constraints~\cite{crooks2007beyond}, namely,
\begin{equation}
\langle S(p)\rangle=\overline{S}, \ \ \ \langle E(p)\rangle = \overline{E},
\end{equation}
yielding a hyperensemble with two Lagrange multipliers $(\alpha,\gamma)$ of the form
\begin{equation}
\label{eq:crooks_hyper}
P_{\gamma,\alpha}(p)=\frac{e^{\alpha S( p)-\gamma E( p)}}{Z(\alpha,\gamma)},
\end{equation}
with $Z(\alpha,\gamma)=\int_{\Delta(X)}e^{\alpha S( p)-\gamma E( p)}d p$. At $\gamma=0$, Eq.\,\eqref{eq:crooks_hyper} becomes the entropic prior (Sec.~\ref{app:entropic}), whereas at $\alpha=0$, it becomes the energetic prior of Sec.~\ref{ssec:expected_average_energy}. Crooks showed that $P(p)$ may be rewritten as depending solely on the Kullback-Leibler divergence of $p$ to a particular reference distribution, namely, the BG equilibrium $\pi^\beta$ at $\beta=\gamma/\alpha$:
\begin{equation}
\label{eq:crooks_hyper_DKL}
P(p)=\frac{e^{-\alpha D( p||\pi^{\beta})}}{\tilde{Z}(\alpha,\gamma)},
\end{equation}
with $\tilde{Z}(\alpha,\gamma)=Z(\alpha,\gamma)Z_{\gamma/\alpha}^{-\alpha}$, where $Z_\beta$ denotes the partition function of $\pi^\beta$. The quantity $\beta^{-1}D(p||\pi^\beta)$ is the excess nonequilibrium free energy $\Delta F_\beta(p):=F_\beta(p)-F_\beta(\pi^\beta)$ (see Sec.~\ref{ssec:BG_comparison}), from which Crooks's hyperensemble may be interpreted as $\mathbb{S}$-maximizing subject to $\langle \Delta F_\beta(p)\rangle=\overline{F}$ for a chosen inverse temperature $\beta$. 

At $\alpha=\infty$, the BG distribution $\pi^\beta$ is recovered from~\eqref{eq:crooks_hyper} with probability $1$. In the limit $\alpha\rightarrow-\infty$, a vanishing entropy $\overline{S}=0$ is achieved, rather than a positive lowest value; that is permitted because the {\it expected} average energy constraint $\langle E(p)\rangle=\overline{E}$ does not require any individual distribution $p$ to exhibit $E(p)=\overline{E}$. Instead, there is a limiting representation in terms of mixtures of Kronecker deltas,
\begin{equation}
\label{eq:Kronecker_mixture}
P_{\gamma,-\infty}(p):=\sum_{x\in X}g_x\delta(p-\delta_x),
\end{equation}
with normalized weights $(g_{x})_{x\in X}$ satisfying $\sum_{x}\epsilon_xg_x=\overline{E}$, i.e., $g\in \Delta_{\overline{E}}$. Any mixture of the form~\eqref{eq:Kronecker_mixture} achieves $\overline{S}=0$ as implied by $\alpha=-\infty$, but it remains to show what the values of $g_x$ are. In this limit, the hyperstatistical uncertainty is over the discrete set $X$, to be quantified via Shannon entropy of $g$. The maximizer of $S(g)$ is thus simply BG with mean energy $\overline{E}$; that is,
\begin{equation}
g_x=\pi^{\beta(\overline{E})}_x=\frac{e^{-\beta(\overline{E})\epsilon_x}}{Z_{\beta(\overline{E})}}.
\end{equation}
If one takes the limit $\alpha\rightarrow-\infty$ at constant $\gamma$, the limiting Kronecker mixture is $g=\pi^\gamma$, i.e., $\beta(\overline{E})=\gamma$. Thus,
\begin{equation}
P_{\gamma,-\infty}(p)=\sum_{x\in X}\frac{e^{-\gamma\epsilon_x}}{Z_{\gamma}}\delta(p-\delta_{x}).
\end{equation}

\section{Hit-and-run sampling}
\label{app:hitandrun}

We provide a hit-and-run algorithm $\mathbf{HR}(\Delta_{E^*})$ that samples from $P_{E^*}$. We assume $\{\epsilon_x\}_{x\in X}$ are all distinct and differ from $E^*\in(\epsilon_-,\epsilon_+)$. We hereafter denote $||v||=\sqrt{\sum_{x\in X}v_x^2}$. In what follows, we describe three parts of the algorithm that require particular attention, (i,ii,iii), and then provide pseudocode.

(i) The initialization step of $\mathbf{HR}(\Delta_{E^*})$ requires construction of an arbitrary distribution in $\Delta_{E^*}$. We achieve this by a uniform mixture of the vertices of $\Delta_{E^*}$. Let $X_-=\{x:\epsilon_x<E^*\}$ and $X_+=\{x:\epsilon_x>E^*\}$; then for any $(x',x'')\in X_-\times X_+$, there is a distribution $p^{(x',x'')}$ with support only on $x'$ and $x''$ residing in $\Delta_{E^*}$ (obtained in Appendix~\ref{app:min_ent}). To ensure full support of the initial point (albeit not strictly necessary~\cite{LovaszSantosh}), we uniformly mix $p^{(x',x'')}$ over all $(x',x'')\in X_-\times X_+$:
\begin{equation}
\label{eq:HR_init}
p^{(1)}=\frac{1}{|X_-||X_+|}\sum_{x'\in X_-}\sum_{x''\in X_+}p^{(x',x'')}.
\end{equation}

(ii) A key step of the algorithm is to generate a randomly oriented vector. In the unconstrained Euclidean space $\mathbb{R}^{|X|}$, this would be achieved by sampling a vector $u$ with i.i.d.\ Gaussian components $\{u_x\}_{x\in X}$. The two hyperplanar constraints require projection into a direction tangent to $\Delta_{E^*}$; we subtract the mean $\frac{1}{|X|}\sum_{x\in X}u_x$ to ensure tangency to $\Delta(X)$, and also project out components normal to $\Delta_{E^*}$ via a mean-zero version of the vector of energies. More concretely, we define $\mathbf{1}:=(1)_{x\in X}$, $\boldsymbol{\epsilon}:=(\epsilon_x)_{x\in X}$, and three unit vectors
\begin{equation}
\label{eq:units}
\hat{1}:=\frac{\mathbf{1}}{\sqrt{n}}, \ \ \ \hat{\epsilon}=\frac{\boldsymbol{\epsilon}}{||\boldsymbol{\epsilon}||}, \ \ \ \hat{e}=\frac{\hat{\epsilon}-(\hat{1}\cdot \hat\epsilon)\hat{1} }{||\hat{\epsilon}-(\hat{1}\cdot \hat\epsilon)\hat{1} ||}.
\end{equation}
Then, $u'=u-(\hat{1}\cdot u)\hat{1}-(\hat{e}\cdot u)\hat{e}$ has a random orientation tangent to $\Delta_{E^*}$.

(iii) Another detail is how to obtain the boundary points of the (randomly oriented) chord passing through a given distribution. Suppose we have a distribution $q=(q_x)_{x\in X}$ and a vector $u'=(u'_x)_{x\in X}$ respecting the normalization and average energy constraints. Since $\Delta_{E^*}$ is a convex polytope whose vertices constitute a measure-zero part of the boundary, a chord through $q$ of the form $\tilde{q}(\zeta)=q+\zeta u'$ will with probability one intersect two faces of $\Delta_{E^*}$. This will occur at the two smallest values of $|\zeta|$ for which $\tilde{q}_x(\zeta)=0$ for some component $x\in X$, once for $\zeta>0$ and once for $\zeta<0$. The condition $\tilde{q}_x=0$ is equivalently
\begin{equation}
\label{eq:tilde_q_0}
q_x+\zeta u'_x=0 \ \Rightarrow \ \zeta = -\frac{q_x}{u'_x}.
\end{equation}
For each $x\in X$, the solution to~\eqref{eq:tilde_q_0} is either positive, $\zeta>0$ (if $u'_x<0$), or negative, $\zeta<0$ (if $u'_x>0$). As $|\zeta|$ grows, for each sign, there will be some $x\in X$ that is reached first---namely, that for which $|q_x/u'_x|$ is minimized. Thus, the positive and negative solutions are
\begin{equation}
\zeta_+=\min_{x:u_x<0}\left\vert\frac{q_x}{u'_x}\right\vert, \ \ \ \zeta_-=-\min_{x:u'_x>0}\left\vert\frac{q_x}{u'_x}\right\vert. 
\end{equation}
All other solutions of~\eqref{eq:tilde_q_0} are outside of $\Delta_{E^*}$, requiring negative probabilities of some components. Having covered the necessary details (i,ii,iii), we below present a full pseudocode for the algorithm $\mathbf{HR}(\Delta_{E^*})$.

\noindent\fcolorbox{black}{white}{\begin{minipage}{8.5cm}
$\mathbf{HR}(\Delta_{E^*})$\ \\

Input arguments: \ $E^*$, \ $\{\epsilon_x\}_{x\in X}$, \ $T$
\begin{enumerate} 
    \item Initialize:
    \begin{itemize}
       \item Compute $p^{(1)}$ by \eqref{eq:HR_init} (with \,\eqref{eq:p_mixture_Estar},\ \eqref{eq:q_mixture_Estar})
        \item Build unit vectors $\hat{1}$ and $\hat{e}$ of \eqref{eq:units}
    \end{itemize}
    \item For $t=1,...,T-1$, repeat:
    \begin{enumerate}
      \item Sample a randomly oriented vector $u'$:
      \begin{itemize}
          \item Sample $u=(u_x)_{x\in X}$ i.i.d.\ $\sim \mathcal{N}(0,1)$
          \item Project as $u'=u-(\hat{1}\cdot u)\hat{1}-(\hat{e}\cdot u)\hat{e}$
    \end{itemize}
    \item Compute intersections of $p^{(t)}+\zeta u'$ with the boundaries of $\Delta_{E^*}$:
        \begin{itemize}
            \item $\zeta_+=\underset{x:u'_x<0}{\min}\, p^{(t)}_x/|u'_x|$
            \item $\zeta_-=-\underset{x:u'_x>0}{\min}\, p^{(t)}_x/u'_x$
        \end{itemize}
        \item Set $p^{(t+1)}=p^{(t)}+\xi\, u'$, where $\xi\sim\mathrm{U}[\zeta_-,\zeta_+]$
    \end{enumerate}
\end{enumerate}
\ \\
Return $p^{(1)},\ldots,p^{(T)}$

\end{minipage}}
\ \\

This HR algorithm can be extended directly to incorporate rejection probabilities at each step. If $p^{(t+1)}$ is rejected (at step (c)), retain the distribution as $p^{(t+1)}=p^{(t)}$ (i.e., take $\xi=0$) and increment $t\rightarrow t+1$. This permits incorporation of biases in the stationary density~\cite{Romeijn_1994}, e.g., the expected Shannon entropy constraint for sampling of $P_{E^*,\alpha}$ (see Appendix~\ref{app:rej_mcmc}).

Additionally, the algorithm is straightforwardly extended to incorporate multiple linear constraints $Q_k(p)=Q^*_k$ for $k\in[K]$ (hyperensemble $\mathrm{U}[\cap_{k\in[K]} \Delta_{Q^*_k}]$), in which case the randomly-oriented vector $u$ is obtained by orthogonal projection. Note also that in the absence of any exact distributional constraints ($K=0$) beyond normalization, HR is still fast-mixing in $\Delta(X)$. The projection step then only ensures normalization; this permits fast sampling of, e.g., $P_\gamma$, $P_\alpha$, and $P_{\gamma,\alpha}$.

\section{Entropy minimizer in $\Delta_{E^*}$}
\label{app:min_ent}

Denoting $\delta_{x}:=(\delta_{x,x'})_{x'\in X}$, the vertices of $\Delta_{E^*}$ are distributions of the form 
\begin{equation}
\label{eq:p_mixture_Estar}
p^{(x',x'')}=q\delta_{x'}+(1-q)\delta_{x''},
\end{equation}
where $\epsilon_{x'}<E^*<\epsilon_{x''}$, with $q=q(x',x'')$ given by
\begin{equation}
\label{eq:q_mixture_Estar}
q(x',x'')=\frac{\epsilon_{x''}-E^*}{\epsilon_{x''}-\epsilon_{x'}}
\end{equation}
from which
\begin{equation}
E\left(p^{(x',x'')}\right)=(\epsilon_{x'}-\epsilon_{x''})q+\epsilon_{x''}=E^*.
\end{equation}
In particular, among all such vertices, the entropy minimizer is $p^{(\hat{x}',\hat{x}'')}$, with
\begin{equation}
\label{eq:min_pair_appendix}
(\hat{x}',\hat{x}''):=\underset{{(x',x''):\epsilon_{x'}<E^*<\epsilon_{x''}}}{\mathrm{argmin}}h(q(x',x'')),
\end{equation}
where $h(q)=-q\log q-(1-q)\log(1-q)$ is the binary entropy function. The minimal value itself is
\begin{equation}
\label{eq:S_star_min}
S_{\mathrm{min}}(E^*):=\underset{p\in\Delta_{E^*}}{\mathrm{min}}S(p)=h\left(q(\hat{x}',\hat{x}'')\right)\le \log 2.
\end{equation}
Although upper bounded by $\log 2$, the actual value of $S_{\mathrm{min}}(E^*)$ depends on how close $q(\hat{x}',\hat{x}'')$ is to either $0$ or $1$. In particular, the entropy-minimizing pair of microstates $(\hat{x}',\hat{x}'')$ is one of two tuples: $(x_-,x_{>})$ or $(x_{<},x_+)$, with $x_{\pm}$ the energy-extremizing microstates and
\begin{equation}
x_{>}=\underset{x:\epsilon_x>E^*}{\mathrm{argmin}}\, \epsilon_{x}, \ \ x_{<}=\underset{x:\epsilon_x<E^*}{\mathrm{argmax}}\, \epsilon_{x}.
\end{equation}
We denote $\epsilon_{>}=\epsilon_{x_{>}}$ and $\epsilon_{<}=\epsilon_{x_{<}}$. Then,
\begin{equation}
\label{eq:s_min_appendix}
S_{\mathrm{min}}(E^*)=\min\left(h\left(\frac{\epsilon_>-E^*}{\epsilon_>-\epsilon_-}\right),h\left(\frac{E^*-\epsilon_<}{\epsilon_{+}-\epsilon_{<}}\right)\right).
\end{equation}
Eq.\,\eqref{eq:s_min_appendix} can be deduced as follows. First, note that entropy is a concave function: for any two distributions $p$ and $p'$, and for any $w\in[0,1]$, the entropy satisfies $S(wp+(1-w)p')\ge wS(p)+(1-w)S(p')$. Consider a set of distributions $K$ and finite subset thereof $\kappa$ for which each distribution in $K$ can be written as a mixture. Then by concavity, the entropy minimizing distribution(s) reside amongst $\kappa$. For $\Delta_{E^*}$ which is a convex polytope, the minimal set $\kappa$ consists of all vertices: mixtures $p^{(x',x'')}$ of Kronecker deltas at $x',x''$ with $\epsilon_{x'}<E^*<\epsilon_{x''}$, with weights $q^{(x',x'')}$ given by~\eqref{eq:q_mixture_Estar}. Among those vertices, the one with mixing weight closest to $0$ or $1$ is either $(x_{<},x_+)$ or $(x_-,x_>)$, and hence by monotonicity of $h(\cdot)$ in both $[0,1/2]$ and $[1/2,1]$, one of those pairs will yield the minimal entropy value across all distributions in $\Delta_{E^*}$.

Note that at any fixed Hamiltonian, the minimum entropy value $S_{\mathrm{min}}(E^*)$ undergoes non-monotonic variation as a function of $E^*$, vanishing whenever $E^*$ crosses one of the energy values. See Fig.~\ref{fig:barS_vs_Estar_alpha}.

\section{$Z(\gamma)$ integral}
\label{app:Zgamma}

To obtain $Z(\gamma)$ of Sec.~\ref{ssec:expected_average_energy}, we use a formula presented in Ref.~\cite{Barvinok1992}. Let $L = \{y \in\mathbb{R}^n:\sum_{j=1}^ny_j= 1\}$ denote the normalization hyperplane in $\mathbb{R}^n$ and let $\tilde{\Delta}=L\cap \mathbb{R}_+^n$. Denote by $\mu$ the
Lebesgue measure on $L$ induced from $\mathbb{R}^n$. Then
\begin{equation}
\int_{\tilde{\Delta}} e^{\langle c,y\rangle} d\mu=\sqrt{n}\sum_{j=1}^n \frac{e^{c_j}}{\prod_{k\in[n]\backslash\{j\}}( c_j - c_k)},
\end{equation}
where $c=(c_1,...,c_n)\in\mathbb{R}^n$, and $\langle c,y\rangle:=\sum_{j=1}^nc_jy_j=y^Tc$. In the setting of $P_{\gamma}$, we have $n=|X|$ and $c_x=-\gamma\epsilon_x$ (or $c=-\gamma\boldsymbol{\epsilon}$ with $\boldsymbol{\epsilon}:=(\epsilon_x)_{x\in X}$). Then with $p=(p_x)_{x\in X}\in\Delta(X)$, we have that $\langle c,y\rangle$ becomes $\langle -\gamma\boldsymbol{\epsilon},p\rangle=-\gamma E(p)$. Finally, note that $d\mu=\sqrt{n}\, dp_1\cdots dp_{n-1}=:\sqrt{n}\, dp$~\cite{knight1966}. Accordingly,
\begin{equation}
\begin{aligned}
Z(\gamma)&=\int_{\Delta(X)} e^{-\gamma E(p)} dp=\frac{1}{\sqrt{n}}\int_{\tilde{\Delta}} e^{-\gamma E(p)} d\mu\\
&=\frac{1}{(-\gamma)^{n-1}}\sum_{x\in X} \frac{e^{-\gamma\epsilon_x}}{\prod_{x'\in X\backslash\{x\}}(\epsilon_x -\epsilon_{x'})}.
\end{aligned}
\end{equation}

\subsection{Linearly spaced energies}
For the linearly spaced energies example $\epsilon_x=\upsilon x$ with sample space $X=\{1,...,n\}$, the hyperensemble $P_\gamma$ (Sec.~\ref{ssec:expected_average_energy}) has a simple form. Namely,
\begin{equation}
P_\gamma(p)=\frac{e^{-\gamma \sum_{x=1}^nxp_x}}{Z(\gamma)}
\end{equation}
with~\eqref{Z_gamma} for $Z(\gamma)$ becoming
\begin{equation}
Z(\gamma)=\frac{1}{(-\upsilon\gamma)^{n-1}}\sum_{x\in X}\frac{e^{-\gamma\upsilon x}}{K_x},
\end{equation}
where 
\begin{equation}
\begin{aligned}
K_x&:=\prod_{x'\in X\backslash\{x\}}(x-x')\\
&=(-1)^{n-x}\Gamma(x)\Gamma(n-x+1).
\end{aligned}
\end{equation}
Thus
\begin{equation}
\begin{aligned}
\label{eq:Z_gamma_level}
Z(\gamma)&=\frac{-1}{(\upsilon\gamma)^{n-1}}\sum_{x=1}^{n}\frac{[-e^{-\gamma\upsilon}]^x}{\Gamma(x)\Gamma(n-x+1)}\\
&=\frac{e^{-\gamma\upsilon}}{(n-1)!}\left(\frac{1-e^{-\gamma\upsilon}}{\gamma\upsilon }\right)^{n-1}.
\end{aligned}
\end{equation}
Using $\overline{E}(\gamma)=-\frac{d}{d\gamma}\log Z(\gamma)$,
\begin{equation}
\label{eq:barE_of_gamma}
\overline{E}(\gamma)=\upsilon+\frac{n-1}{\gamma}\left[\frac{\gamma\upsilon}{1-e^{\gamma\upsilon}}+1\right].
\end{equation}
The natural parameter matching $E(\pi^\beta)=\overline{E}(\gamma)$ at $\upsilon=1$ then reduces to~\eqref{eq:transcendental_gamma_beta}.

\section{Elementary rejection algorithms}
\label{app:rej_mcmc}

We here describe rejection-based sampling algorithms for the hyperensembles considered in this work. Each is as follows: we have a prescribed hyperensemble $P(p)$, and use rejection sampling on candidates from $\mathrm{Dirichlet}(\mathbf{1})$ (see Appendix~\ref{app:dirichlet_entropic_crooks}), or $\mathrm{U}[\Delta_{Q^*}]$ for general sharp constraints (Eq.\,\eqref{eq:Qk}), accepting $p^{(1)}$ with probability $\propto P(p)$. Then, an MCMC proceeds via a Markov chain whose stationary density is uniform on $\Delta(X)$ (or $\Delta_{Q^*}$) is used to sample candidate transitions, $p^{(t)}\rightarrow p'$, with $p^{(t+1)}=p'$ if accepted and otherwise $p^{(t+1)}=p^{(t)}$. The initialization of $p^{(1)}$ via rejection sampling makes this algorithm start off being statistically exact already, with all subsequent samples having the correct marginal distribution, avoiding a transient.

\subsection{Expectation constraints}
We start with samplers for hyperensembles $P$ with full support over $\Delta(X)$, with our main focus being on $P_\gamma$ (the $\mathbb{S}$-maximizer subject to $\langle E(p)\rangle=\overline{E}$), as well as the posterior hyperensemble under observation of $\mathbf{x}$ under Crooks's prior.

A simple rejection algorithm for $P_{\gamma}$ is as follows: sample a candidate $p\sim \mathrm{Dirichlet}(\mathbf{1})$, then to accept $p$ with probability $\mathcal{A}_\gamma(p)\propto e^{-\gamma E(p)}$. With the optimal choice of scalar multiple, the acceptance probability is
\begin{equation}
\mathcal{A}_\gamma(p)=\begin{cases}
e^{-\gamma(E(p)-\epsilon_-)}, & \gamma\ge 0\\
e^{-\gamma(E(p)-\epsilon_+)}, & \gamma<0\\
\end{cases},
\end{equation}
ensuring $\mathcal{A}_\gamma(p)\le 1$ for all $p\in\Delta(X)$ and that $\mathcal{A}_\gamma(p)=1$ is attained for the mode---namely, $A_\gamma(\delta_{x_+})=1$ for $\gamma<0$ and $\mathcal{A}_{\gamma}(\delta_{x_-})=1$ for $\gamma\ge 0$, with $\delta_x=(\delta_{x,x'})_{x'\in X}$ denoting the Kronecker delta distribution at $x$. The mean value of $\mathcal{A}_{\gamma}(p)$ is $\Theta(1)$ for $|\epsilon_x|=O(1)$ and $\gamma=O(1)$. After obtaining a successfully accepted candidate, one may proceed with MCMC under acceptance probability $\mathcal{A}_\gamma(p\rightarrow p')=\min(1,e^{-\gamma[E(p')-E(p)]})$ of $p\rightarrow p'$.

The same considerations apply to sample from $P_{\alpha}$ via Dirichlet rejection, accepting a candidate $p$ with probability proportional to $P_{\alpha}(p)$. For $\alpha>0$, the mode is the uniform on $X$ (entropy $\log n$); for $\alpha<0$, all $n:=|X|$ point masses $\delta_{x}$ (entropy zero), $x\in X$, are tied as the mode. 
Taking a candidate $p\sim \mathrm{Dirichlet}(\mathbf{1})$; in particular, if $\alpha\ge 0$, accept with probability
\begin{equation}
\mathcal{A}_{\alpha}(p)=e^{\alpha [S(p)-\log n]},
\end{equation}
with multiplicative constant $e^{-\alpha\log n}=1/n^\alpha$ ensuring $\mathcal{A}_\alpha(p)\le 1$ for all $p\in\Delta(X)$, with equality only at the mode. If $\alpha<0$, lower entropy values are favored, with $\mathcal{A}_{\alpha}(p)$ reaching one only at the modes $(\delta_{x})_{x\in X}$ where entropy vanishes. In that case the optimal acceptance probability has multiplicative constant $1=e^{\alpha S(\delta_x)}$ giving
\begin{equation}
\mathcal{A}_{\alpha}(p)=e^{\alpha S(p)}.
\end{equation}
Any number of MCMC steps following an initially accepted candidate can be applied, following a similar protocol to that of $P_\gamma$ but with $(-\gamma,E(p))\rightarrow (\alpha,S(p))$. In particular, the transition $p\rightarrow p'$ is accepted with probability $\mathcal{A}_{\alpha}(p\rightarrow p')=\min(1,e^{\alpha[S(p')-S(p)]})$.

Identical procedures hold for Crooks's hyperensemble $P_{\gamma,\alpha}(p)$ with $\mathcal{A}_{\gamma,\alpha}(p)\propto e^{-\gamma\Delta F_{\beta}(p)}$ at $\beta=\gamma/\alpha$. The mode for $\alpha>0$ is $\pi^\beta$ with $F_\beta(\pi^\beta)=-\beta^{-1}\log Z_\beta$, and for $\alpha<0$ it is $\delta_{x_+}$ (if $\gamma>0$) for which
\begin{equation}
F_\beta(\delta_{x_+})=\max_{p\in\Delta(X)}F_{\beta}(p)=\epsilon_+,
\end{equation}
and likewise for $x_-$ if $\gamma<0$. Transition $p\rightarrow p'$ is accepted with probability
\begin{equation}
\begin{aligned}
\mathcal{A}_{\gamma,\alpha}(p\rightarrow & p')=\min(1,e^{-\gamma(F_\beta(p')-F_\beta(p))})\\
&=\min(1,e^{-\gamma[E(p')-E(p)]+\alpha[S(p')-S(p)]}).
\end{aligned}
\end{equation}
These considerations also apply to $P_{\gamma,\alpha}(\cdot|\mathbf{x})$, the posterior hyperensemble upon observation of microstate sequence $\mathbf{x}$ under Crooks's prior. After an initial rejection-sampling step, MCMC proceeds with candidate transition $p\rightarrow p'$ accepted with probability $\mathcal{A}(p\rightarrow p')=\min(1,e^{\mathcal{H}(p,p')})$ where
\begin{equation}
  \mathcal{H}(p,p'):=-\gamma( F_{\beta}(p')-F_{\beta}(p))+\sum_{x\in X}N_x\log \frac{p'_x}{p_x},
\end{equation}
with $\beta=\gamma/\alpha$ as above.

\subsection{Exact constraints}

We may also draw $p\sim P_{E^*,\alpha}$ via rejection sampling. Unlike the aforementioned models, $P_{E^*,\alpha}$ has a restricted support, $\Delta_{E^*}$. But we have already obtained in Sec.~\ref{ssec:exact_average_energy} an efficient hit-and-run algorithm for uniform sampling of $\Delta_{E^*}$, which enables this approach for $P_{E^*,\alpha}$. In particular, an initial distribution is drawn from $P_{E^*}$ and accepted with probability $\mathcal{A}_{E^*,\alpha}(p)\propto e^{\alpha S(p)}$; with optimal coefficients, the acceptance probability is
\begin{equation}
\mathcal{A}_{E^*,\alpha}(p)=\begin{cases}
e^{\alpha(S(p)-S_{\mathrm{max}}(E^*))}, & \ \alpha\ge 0\\
e^{\alpha(S(p)-S_{\mathrm{min}}(E^*))}, & \ \alpha<0
\end{cases}
\end{equation}
where $S_{\mathrm{max}}(E^*)=S(\pi^\beta)$ and $S_{\mathrm{min}}(E^*)$ was characterized in Sec.~\ref{ssec:exact_average_energy} (and Appendix~\ref{app:min_ent}). For subsequent MCMC steps, candidate transition $p\rightarrow p'$ is obtained via a step of $\mathbf{HR}(\Delta_{E^*})$, and accepted with probability $\min(1,e^{\alpha [S(p')-S(p)]})$.

\section{Limiting regimes of excess free energy}
\label{app:beta0_KLvals}

We consider $D(q||\pi^\beta)$ for $q$ taken as random from, or as the mean of, a hyperensemble $P$. Specifically, we characterize limiting cases of the following:
\begin{equation}
\begin{aligned}
\label{eq:KL_defs_app}
D_{\gamma}(\overline{p})&:=D(\langle p\rangle_{P_{\gamma(\beta)}}||\pi^\beta), \\ 
D_{E^*}(\overline{p})&:=D(\langle p\rangle_{P_{E^*(\beta)}}||\pi^\beta), \\
\overline{D_{\gamma}(p)}&:=\langle D(p||\pi^\beta)\rangle_{P_{\gamma(\beta)}}, \\ 
\overline{D_{E^*}(p)}&:=\langle D(p||\pi^\beta)\rangle_{P_{E^*(\beta)}}.
\end{aligned}
\end{equation}
We for simplicity consider $\beta\ge 0$, with $\beta<0$ having mirrored properties. (E.g., the $\beta\rightarrow\infty$ considerations presented below hold also for $\beta\rightarrow-\infty$, under replacement of $(x_-,\epsilon_-)$ by $(x_+,\epsilon_+)$.)

\subsection{$\beta\rightarrow\infty$ limit}

The first special case of interest is $\beta\rightarrow\infty$, where, across models, the probability concentrates onto the microstate $x_-$ with minimal energy $\epsilon_-$. The BG equilibrium becomes a Kronecker delta, $\lim_{\beta \rightarrow\infty}\pi^\beta=\delta_{x_-}$. The $\beta\rightarrow\infty$ limit corresponds to $\gamma\rightarrow\infty$ and to $E^*\rightarrow\epsilon_-$, so the hyperensembles $P_\gamma$ and $P_{E^*}$ both concentrate:
\begin{equation}
\lim_{E^*\rightarrow\epsilon_-}P_{E^*}(p)=\lim_{\gamma \rightarrow\infty}P_{\gamma}(p)=\delta(p-\delta_{x_{-}}),
\end{equation}
from which $D(p||\pi^\beta)=0$ with probability $1$. The mean distributions then also trivially become $\langle p_x\rangle=\delta_{x,x_-}$, so that $D(\overline{p}||\pi^\beta)$ also vanishes. Hence, all quantities of~\eqref{eq:KL_defs_app} vanish in this limit, as reflected in Fig.~\ref{fig:KL}.

\subsection{KL divergences as $\beta\rightarrow 0$}
In the $\beta\rightarrow 0$ limit (infinite temperature), BG becomes uniform: $\lim_{\beta\rightarrow 0}\pi^\beta_x=\pi^0_x=1/n$. This corresponds in $P_\gamma$ to $\gamma\rightarrow 0$, and in $P_{E^*}$ to $E^*\rightarrow \mathcal{E}$ (the arithmetic average energy, i.e., the average energy of the uniform distribution; $E(\pi^0)=\mathcal{E}$).

Only one of the KL divergence properties of~\eqref{eq:KL_defs_app} vanishes at $\beta=0$, namely, $D_{\gamma}(\bar{p})$, because $P_\gamma$ approaches $\mathrm{Dirichlet}(\mathbf{1})$ and thus
\begin{equation}
\lim_{\gamma\rightarrow 0}\langle p_x\rangle_{P_{\gamma}}=\langle p_x\rangle_{\mathrm{U}[\Delta(X)]}=\frac{1}{n},
\end{equation}
so $D_\gamma(\overline{p})\rightarrow 0$ as $\beta\rightarrow 0$, as is consistent with Fig.~\ref{fig:KL}. In contrast, the other three quantities in~\eqref{eq:KL_defs_app} have nontrivial limiting values at $\beta=0$, i.e., predict excess nonequilibrium free energy even at infinite temperature. We consider each of the remaining cases below.

\subsubsection{Mean KL divergence in $P_\gamma$ at $\beta=0$}
Considering $P_\gamma$ as $\gamma\rightarrow 0$, the mean KL divergence behaves as
\begin{equation}
\begin{aligned}
\label{eq:limit_gamma0_meanD}
\lim_{\gamma\rightarrow 0}\overline{D_\gamma(p)}&=\langle D(p||\pi^0)\rangle_{\mathrm{U}[\Delta(X)]}\\
&=\log n-\langle S(p)\rangle_{\mathrm{U}[\Delta(X)]}.
\end{aligned}
\end{equation}
The expected entropy of a Dirichlet distribution is known \cite{wolpert1995,nemenman2001}; for the uniform case, it has value
\begin{equation}
\label{eq:limit_gamma0_meanD_2}
\langle S(p)\rangle_{\mathrm{U}[\Delta(X)]}=\psi_0(n+1)-1+\gamma_E,
\end{equation}
with $\gamma_E\approx 0.577$ the Euler-Mascheroni constant and $\psi_0(z)$ the digamma function; for integer argument $z$, $\psi_0(z)=-\gamma_E+\sum_{k=1}^{z-1}\frac{1}{k}$, so~\eqref{eq:limit_gamma0_meanD_2} becomes
\begin{equation}
\label{eq:bar_D_gamma_0}
\lim_{\gamma\rightarrow 0}\overline{D_\gamma(p)}=1+\log n-H_{n},
\end{equation}
with $H_n:=\sum_{k=1}^{n}\frac{1}{k}$ the $n$th Harmonic number. Eq.\,\eqref{eq:bar_D_gamma_0} has large-$n$ asymptotic $1-\gamma_E-\frac{1}{2n}$.

\subsubsection{$\beta\rightarrow 0$ limiting values in $P_{E^*}$}

As $\beta\rightarrow 0$ the hyperenesmble $P_{E^*}$ approaches $P_{\mathcal{E}}$, the uniform density on the energy hyperplane passing through the uniform distribution $\pi^0$. The limiting mean KL divergence is
\begin{equation}
\begin{aligned}
\label{eq:bar_D_Estar_mathcalE}
\lim_{E^*\rightarrow\mathcal{E}}\overline{D_{E^*}(p)}&=\langle D(p||\pi^0)\rangle_{P_{\mathcal{E}}}\\
&=\log n-\frac{1}{|\Delta_{\mathcal{E}}|}\int_{\Delta_{\mathcal{E}}}\, S(p)\, dp,
\end{aligned}
\end{equation}
requiring the average entropy in the uniform density over the energy surface $\Delta_{\mathcal{E}}$. The entropy averaging method of Ref.~\cite{wolpert1995} does not directly apply to compute~\eqref{eq:bar_D_Estar_mathcalE}, but its value can be estimated by hit-and-run sampling (Appendix~\ref{app:hitandrun}).

The $\beta\rightarrow 0$ limit of the KL divergence of the mean distribution to $\pi^\beta$ is
\begin{equation}
\begin{aligned}
\label{eq:D_Estar_mathcalE}
\lim_{E^*\rightarrow\mathcal{E}}D_{E^*}(\overline{p})&=D(\langle p \rangle_{P_{\mathcal{E}}}||\pi^0)\\
&=\log n-S(\mathrm{centroid}(\Delta_{\mathcal{E}})).
\end{aligned}
\end{equation}
The centroid of the polytope $\Delta_{\mathcal{E}}$ can be computed via triangulation~\cite{de_Berg_2008} or estimated by hit-and-run sampling (Appendix~\ref{app:hitandrun}). 

In simple cases, direct analytical evaluation of~\eqref{eq:bar_D_Estar_mathcalE} and~\eqref{eq:D_Estar_mathcalE} is possible; for $n=3$ and $\epsilon_x=x$, we have that $\mathcal{E}=2$ and
\begin{equation}
    \Delta_{\mathcal{E}}=\left\{\left(\frac{1-\phi}{2},\phi,\frac{1-\phi}{2}\right)\right\}_{\phi\in[0,1]},
\end{equation}
and the centroid is $(\frac{1}{4},\frac{1}{2},\frac{1}{4})$, from which~\eqref{eq:D_Estar_mathcalE} evaluates to $\log\frac{3}{2\sqrt{2}}=0.058\, 891\, \ldots$ and~\eqref{eq:bar_D_Estar_mathcalE} evaluates, via integration with respect to $\phi$, to $\log\frac{3}{\sqrt{2}}-\frac{1}{2}=0.252\, 039\, \ldots$, and these both match Fig.~\ref{fig:KL}.

\section{Aligning $f(\beta)$ with $P_\gamma$}
\label{app:f_beta_for_P_gamma}

In Sec.~\ref{ssec:P_Estar_mixture_f_beta}, we showed that the hyperensemble approach is compatible with general superstatistics models by constructing a mixture model of $P_{E^*}$ associated with an arbitrary density $f(\beta)$. We now consider the special case of comparison of superstatistics to $P_\gamma$ (Sec.~\ref{ssec:expected_average_energy}), the $\mathbb{S}$-maximizer subject to expectation constraint $\langle E(p)\rangle=\overline{E}$. A wide range of superstatistics models are compatible with that form of energetic constraint; it implies a condition on the density of average energy $\mu$ (or equivalently $f$) of the form
\begin{equation}
\overline{E}=\int_{\epsilon_-}^{\epsilon_+} \mu(E^*)E^*dE^*=\int_{\mathbb{R}} f(\beta)E(\pi^\beta)d\beta.
\end{equation}
To maximally align a superstatistics model with $P_\gamma$, we choose $f(\beta)$ so that the probability density of average energy $\mu(E^*)$ equals that of $P_\gamma$, the latter denoted $\mu_{\gamma}(E^*)$. Since $P_\gamma(p)$ depends on $p$ only through $E(p)$, the density of average energy is
\begin{equation}
\label{eq:mu_gamma_Estar}
\mu_\gamma(E^*)=\int_{\Delta_{E^*}}P_\gamma(p)dp=\frac{|\Delta_{E^*}|e^{-\gamma E^*}}{Z(\gamma)},
\end{equation}
determining the only compatible $f(\beta)=f_\gamma(\beta)$. The superstatistics model for which $\mu(E^*)=\mu_\gamma(E^*)$ can be seen as a quasi-equilibrium approximation of $P_\gamma$. $f_\gamma$ is obtained from~\eqref{eq:mu_gamma_Estar} by a transformation of random variables:
\begin{equation}
\label{eq:f_beta_from_mu_Estar}
f_\gamma(\beta)=\frac{1}{Z(\gamma)}\frac{d^2 \log Z_\beta}{d\beta^2}|\Delta_{E^*(\beta)}|e^{-\gamma E^*(\beta)},
\end{equation}
having used $E^*(\beta)=-\frac{d}{d\beta}\log Z_\beta$ for the BG partition function $Z_\beta$. Note that the second derivative in~\eqref{eq:f_beta_from_mu_Estar} is the energy variance at equilibrium,~\eqref{eq:energy_variance}. Note that the predicted support of $\beta$ is $\mathbb{R}$ rather than $\mathbb{R}_+$, i.e., samples $\pi^\beta$ with $\beta<0$ will arise in the $f_\gamma$-superstatistics model. An elementary direct approach to sample $\beta$ from~\eqref{eq:f_beta_from_mu_Estar} is to draw $p\sim P_\gamma$ and set $\beta=\beta(E(p))$. Examining the $f_\gamma$-superstatistics model alongside $P_\gamma$ allows comparison of quasi-equilibrium and fully nonequilibrium settings. In particular, it does so with aligned density of average energy, with that density being the one arising from $\mathbb{S}$-maximization subject to $\langle E(p)\rangle=\overline{E}$.

\section{Free energy of posterior mode}
\label{app:mode_F}
Recall from Sec.~\ref{ssec:inference_crooks} that $p^*=\tilde{\pi}^{\beta}:=\frac{1}{z}e^{-\beta\tilde{\epsilon}_x}$ with $\tilde{\epsilon}_x=\epsilon_x-\beta^{-1}W(\frac{z}{\alpha}e^{\beta\epsilon_x}N_x)$, with $\beta=\gamma/\alpha$ and $\beta>0$ assumed. Therefore, note that
\begin{equation}
\begin{aligned}
D(p^*||\pi^{\beta})&=\sum_{x\in X}\tilde{\pi}^{\beta}_x\log\frac{\tilde{\pi}^{\beta}_x}{\pi^{\beta}_x}\\
=\log\frac{Z_{\beta}}{z}&+\frac{1}{z}\sum_{x\in X}e^{-\beta\epsilon_x+W(\frac{z}{\alpha}e^{\beta\epsilon_x}N_x)}W(\frac{z}{\alpha}e^{\beta\epsilon_x}N_x)\\
\end{aligned}
\end{equation}
Using $W(y)e^{W(y)}=y$,
\begin{equation}
\begin{aligned}
D(p^*||\pi^{\beta})&=\log\frac{Z_{\beta}}{z}+\frac{1}{z}\sum_{x\in X}e^{-\beta\epsilon_x}\frac{z}{\alpha}e^{\beta\epsilon_x}N_x\\
&=\log\frac{Z_{\beta}}{z}+\frac{1}{\alpha}\sum_{x\in X}N_x=\log\frac{Z_{\beta}}{z}+\frac{N}{\alpha}.
\end{aligned}
\end{equation}
Therefore the free energy of $p^*$ with respect to the reference equilibrium $\pi^{\beta}$ is
\begin{equation}
\begin{aligned}
\Delta F_{\beta}(p^*)&=\beta^{-1}D(p^*||\pi^{\beta})\\
&=\beta^{-1}\log Z_{\beta}+\frac{N-\alpha\log z}{\gamma}
\end{aligned}
\end{equation}
Finally, noting that the equilibrium free energy is $F_{\beta}(\pi^{\beta})=-\beta^{-1}\log Z_{\beta}$, we have that the nonequilibrium free energy of $p^*$ is
\begin{equation}
\begin{aligned}
\label{eq:F_pstar}
F_{\beta}(p^*)=\Delta F_{\beta}(p^*)+F_{\beta}(\pi^{\beta})=\frac{N-\alpha\log z}{\gamma}.
\end{aligned}
\end{equation}
Note that for $N=0$, when $z=Z_{\beta}$, Eq.\,\ref{eq:F_pstar} reduces to the value $F_{\beta}(\pi^{\beta})$, reflecting that $p^*=\pi^{\beta}$ in the absence of data.

\end{document}